\documentclass[prc,twocolumn,preprintnumbers,showpacs]{revtex4}
\usepackage{graphicx,latexsym,amssymb}
\usepackage{dcolumn}
\usepackage{bm}
\usepackage{longtable}%
\begin{document}
\newcommand{\Expt}  {Ex\-peri\-ment}
\newcommand{\Expts} {Ex\-peri\-ments}
\newcommand{\Expl}  {Ex\-peri\-men\-tal}
\newcommand{\expt}  {ex\-peri\-ment}
\newcommand{\expts} {ex\-peri\-ments}
\newcommand{\expl}  {ex\-peri\-men\-tal}
\newcommand{\spcF}  {spectroscopic factor}
\newcommand{\spcFs} {spectroscopic factors}
\newcommand{\sglp}  {single-par\-ticle}
\newcommand{\sglh}  {single-hole}
\newcommand{\ph}    {par\-ticle-hole}
\newcommand{\Ph}    {Par\-ticle-hole}
\newcommand{\coef}  {coefficient}
\newcommand{\coefs} {coefficients}
\newcommand{\Cfg}   {Con\-figu\-ration}
\newcommand{\cfg}   {con\-figu\-ration}
\newcommand{\cfgs}  {con\-figu\-rations}
\newcommand{\ONrule}{ortho-normality rule}
\newcommand{\angD}  {angular dis\-tri\-bution}
\newcommand{\angDs} {angular dis\-tri\-butions}
\newcommand{\AngDs} {Angular dis\-tri\-butions}
\newcommand{\excF}  {excitation function}
\newcommand{\excFs}  {excitation functions}
\newcommand{\ExcFs}  {Excitation functions}
\newcommand{\peneTra} {pene\-tra\-bi\-li\-ty}
\newcommand{\peneTras} {pene\-tra\-bi\-li\-ties}
\newcommand{\Cx} {\mbox{$^{12}$C}}
\newcommand{\Cy} {\mbox{$^{13}$C}}
\newcommand{\Cz} {\mbox{$^{14}$C}}
\newcommand{\Nx} {\mbox{$^{14}$N}}
\newcommand{\Ny} {\mbox{$^{15}$N}}
\newcommand{\Ox} {\mbox{$^{16}$O}}
\newcommand{\Oy} {\mbox{$^{17}$O}}
\newcommand{\Nax} {\mbox{$^{23}$Na}}
\newcommand{\ClF} {\mbox{$^{35}$Cl}}
\newcommand{\ClS} {\mbox{$^{37}$Cl}}
\newcommand{\Ce} {\mbox{$^{140}$Ce}}
\newcommand{\Sm} {\mbox{$^{147}$Sm}}
\newcommand{\Bi} {\mbox{$^{209}$Bi}}
\newcommand{\BiE} {\mbox{$^{208}$Bi}}   
\newcommand{\BiS} {\mbox{$^{207}$Bi}}   
\newcommand{\Pb} {\mbox{$^{208}$Pb}}
\newcommand{\PbF} {\mbox{$^{204}$Pb}}   
\newcommand{\PbX} {\mbox{$^{206}$Pb}}   
\newcommand{\PbS} {\mbox{$^{207}$Pb}}   
\newcommand{\PbN} {\mbox{$^{209}$Pb}}   
\newcommand{\PbT} {\mbox{$^{210}$Pb}}   
\newcommand{\Tl} {\mbox{$^{207}$Tl}}
\newcommand{\TlE} {\mbox{$^{208}$Tl}}   
\newcommand{\Ex}  {E\mbox{${_x}$}}	
\newcommand{\Ep}  {E\mbox{${_p}$}}	
\newcommand{\gam}  {\mbox{$\gamma$}}
\newcommand{\bet}  {\mbox{$\beta$}}
\newcommand{\bgam}  {\mbox{$\beta$--$\gamma$}}
\newcommand{\dpgam}  {\mbox{(d, p$\gamma$)}}
\newcommand{\ppgam}  {\mbox{(p, p'$\gamma$)}}
\newcommand{\ggam}  {\mbox{$\gamma$--$\gamma$}}
\newcommand{\pps}     {(p,p{\mbox{$'$)}}}
\newcommand{\talpha}  {(t,{\mbox{$\alpha$)}}}
\newcommand{\dHe}  {(d, {\mbox{$^3$He)}}}
\newcommand{\sOhlb} {s{\mbox{$_{{1}/{2}}$}}}   
\newcommand{\pOhlb} {p{\mbox{$_{{1}/{2}}$}}}   
\newcommand{\pThlb} {p{\mbox{$_{{3}/{2}}$}}}   
\newcommand{\dThlb} {d{\mbox{$_{{3}/{2}}$}}}   
\newcommand{\dFhlb} {d{\mbox{$_{{5}/{2}}$}}}   
\newcommand{\fFhlb} {f{\mbox{$_{{5}/{2}}$}}}   
\newcommand{\fShlb} {f{\mbox{$_{{7}/{2}}$}}}   
\newcommand{\gShlb} {g{\mbox{$_{{7}/{2}}$}}}   
\newcommand{\gNhlb} {g{\mbox{$_{{9}/{2}}$}}}   
\newcommand{\hNhlb} {h{\mbox{$_{{9}/{2}}$}}}   
\newcommand{\hEhlb} {h{\mbox{$_{{11}/{2}}$}}}  
\newcommand{\iEhlb} {i{\mbox{$_{{11}/{2}}$}}}  
\newcommand{\iThlb} {i{\mbox{$_{{13}/{2}}$}}}  
\newcommand{\jThlb} {j{\mbox{$_{{13}/{2}}$}}}  
\newcommand{\jFhlb} {j{\mbox{$_{{15}/{2}}$}}}  
\newcommand{\Ohlb}  {\mbox{${\frac{1}{2}}$}}       
\newcommand{\Thlb}  {\mbox{${\frac{10}{2}}$}}      
%
%
\newcommand{\apprsim} {\mbox{${< \atop \sim}$}}	
\title
{
The \iEhlb\fFhlb\ and \iEhlb\pThlb\ neutron \ph\ multiplets
~in \Pb
}
\author{A. Heusler
}
\affiliation{%
Max-Planck-Institut f\"ur Kernphysik, 
D-69029 Heidelberg, Germany
}%
\email[correspondance to: ]{A.Heusler@mpi-hd.mpg.de}
\author{
G. Graw
}
\affiliation{%
Sektion Physik der Universit\"at M\"unchen, 
D-85748 Garching, Germany
}
\thanks{supported by MLL and DFG C4-Gr894/2-3}
\author{
R. Hertenberger
}
\affiliation{%
Sektion Physik der Universit\"at M\"unchen, 
D-85748 Garching, Germany
}
\thanks{supported by MLL and DFG C4-Gr894/2-3}
\author{
F. Riess
}
\affiliation{%
Sektion Physik der Universit\"at M\"unchen, 
D-85748 Garching, Germany
}
\thanks{supported by MLL and DFG C4-Gr894/2-3}
\author{
H.-F. Wirth
}
\altaffiliation[now ]{
Physik Department E18,
Technische Universit\"at M\"unchen, 
D-85748 Garching, Germany}
\affiliation{%
Sektion Physik der Universit\"at M\"unchen, 
D-85748 Garching, Germany
}
\thanks{supported by MLL and DFG C4-Gr894/2-3}
\author{%
T. Faestermann
}
\affiliation{Physik Department E12, Technische Universit\"at M\"unchen, 
D-85748 Garching, Germany}
\author{%
R. Kr\"ucken,
}
\affiliation{Physik Department E12, Technische Universit\"at M\"unchen, 
D-85748 Garching, Germany}
\author{
J. Jolie
}
\affiliation{%
Institut f\"ur Kernphysik,
Universit\"at zu K\"oln, 
D-50937 K\"oln, Germany
}
\author{
N. Pietralla
}
\affiliation{%
Institut f\"ur Kernphysik,
Universit\"at zu K\"oln, 
D-50937 K\"oln, Germany
}
\author{
P. von Brentano
}
\altaffiliation[supported by DFG BR799/12-1 ]{~}
\affiliation{%
Institut f\"ur Kernphysik,
Universit\"at zu K\"oln, 
D-50937 K\"oln, Germany
}
%
\begin{abstract}
Inelastic proton scattering via isobaric analog resonances allows to
derive rather complete information about neutron particle-hole states.
We applied this method to the doubly-magic nucleus \Pb\ by measuring
\angDs\ of \Pb(p,~p') on top of the   isobaric analog resonances 
in \Bi\ with the Q3D magnetic spectrograph 
at M\"unchen.
We identify the six states of the \iEhlb\fFhlb\ multiplet and the four
states of the \iEhlb\pThlb\ multiplet in the energy range
$4.6\,MeV<E_x<5.3$\,MeV.  Firm spin assignments for the ten states are
given, some of them new.
Additional measurements of the reaction \PbS(d,~p) confirm the
fragmented \iEhlb\pOhlb\ multiplet.
%
\end{abstract}
\pacs{
21.10.-k,
21.10.Hw,
21.10.Jx,
21.60.-n,
21.60.Cs,
25.40.Ep,
25.40.Ny
}
\maketitle
\section{Introduction
}

\Pb\ is the heaviest easily accessible doubly magic nucleus.
It is an ideal test laboratory to study the shell model in detail. 
For a wide range in excitation energy one-particle one-hole
excitations are dominant, only at an excitation energy above
$E_x$=5.3\,MeV four quasiparticle excitations resulting from
collective two-phonon octupole modes \cite{Yates96c,Valn2001}, start
to contribute.
The structure of the observed states is in agreement with theoretical
expectations up to $E_x$=4.5\,MeV \cite{Rad1996,Schr1997,Valn2001}.
At higher energies, despite impressive experimental research
\cite{NDS1971,NDS1986} not all states expected from the shell model
have been detected. 
Many spin assignments are still ambiguous and in addition there are
more states than expected in the 1-particle 1-hole frame.

Much information has been obtained by inelasting proton scattering via
isobaric analog resonances (IAR-pp').
IAR-pp' is a selective reaction, sensitive to the neutron
particle-hole components of the structure only.  In this way the
observed cross sections, its \excFs\ and \angDs, provide direct
information on quantum numbers and amplitudes of the respective \ph\ 
\cfgs.

Early measurements of inelastic proton scattering via IAR
\cite{1967MO25,1967RI13,1968BO16,1968CR05,1968VO02,1968WH02,
Zai1968,1969RI10,1970KU13} provided detailed information about the
complex mixture of the neutron \ph\ \cfgs.
Since many years little work has been done in this field, we mention,
however, the recent work done with the EUROBALL cluster detector
\cite{Rad1996}. 
The main reason is the need for an energy re\-so\-lu\-tion of better
than 5\,keV in the spectra.  At excitation energies $E_x$=4.2-4.8\,MeV
in \Pb\ there are a few doublets with a spacing below 10\,keV. 
In the region $E_x$=5-6\,MeV the average distance of the known levels
is already less than 10\,keV.  The work done in the 1960s by
\cite{Zai1968,1968WH02,1969RI10,1970KU13} improved the energy
re\-so\-lu\-tion from 35 to 18\,keV; some data were obtained using a
magnetic spectrograph of 9-13\,keV re\-so\-lu\-tion
\cite{1967MO25,1969RI10}.

The present status of the Q3D facility  at M\"unchen
\cite{LMUrep2000p70,LMUrep2000p71,Wirth1999,RalfsQ2005} allows 
to take (p,~p') spectra with a re\-so\-lu\-tion of about 3\,keV within
an energy span of around 1\,MeV and high counting statistics on all
known IAR in \Bi\ and up to excitation energies of at least 8\,MeV.

The IAR-pp' data are complemented by high statistics \PbS(d,~p)
neutron transfer spectra, where the observed transfer quantum numbers
are identical with the neutron particles coupled to the \pOhlb\ hole
configuration of the target \PbS\ in its ground state.

In this paper we concentrate on the energy range $E_x$=4.5 -5.3\,MeV,
a region of considerable level density (at least 35 levels).  We
identify  all members of the shell model \cfgs\ \iEhlb\fFhlb\ and
\iEhlb\pThlb. 
Because of the large value of the orbital angular momentum, the
multiplet states based on the \iEhlb\ neutron particle are weakly
excited.  It is one further step towards the goal of complete
spectroscopy, as started in the early attempt \cite{AB1973} to derive
``complete wave functions'' and the residual interaction among 
1-particle 1-hole
\ph\  \cfgs\ from \expt al data alone.

\section{Shell model
}

\begin{figure}[htb]
\caption[
IAR.scenario
]
{\label{IAR.scenario}%
Sketch of the IAR-pp' scenario for \Pb(p,~p')$\Pb^{*}$ (scale of
proton energy $E_p,E_{p'}$ at left).
A single IAR state with spin $LJ$=\iEhlb\ as the second member of
the isobaric analog multiplet $[\PbN,\Bi,\cdots]$  is exemplified with
{\it one} \cfg\ \iEhlb\fFhlb, but all 45 
excess neutrons  $lj,\nu$ including $\iEhlb$ participate equally, see 
Eq.~\ref{eq.IAR.sum}. 
The \excFs\ of all IAR  are shown \cite{1968WH02}; for the two weakest
IAR \iEhlb, \jFhlb\ they are barely visible, therefore
they are enhanced by a factor~10 (thick curves  at left).
The energies of the particle and hole \cfgs\ $|LJ,\nu>$, $|lj,\nu>$
are taken from \cite{NDS1986}. 
The \peneTra\ of the Coulomb barrier can be estimated from the
comparison of the  maxima for the \gNhlb, \gShlb\ (drawn) and the
\dFhlb, \dThlb\ (dotted) IAR.
Similarly the \peneTra\ of the outgoing particles $lj,\nu$ can be seen
from the comparison of the mean cross section for the \ph\ \cfgs\
$|\iEhlb,\nu>\otimes|lj,\nu>$ with spins $I=J-j,\dots,j+j$
calculated from the s.p. widths derived by this work (lower left).
}
\resizebox{\hsize}{12.50cm}{
{\includegraphics[angle=00]{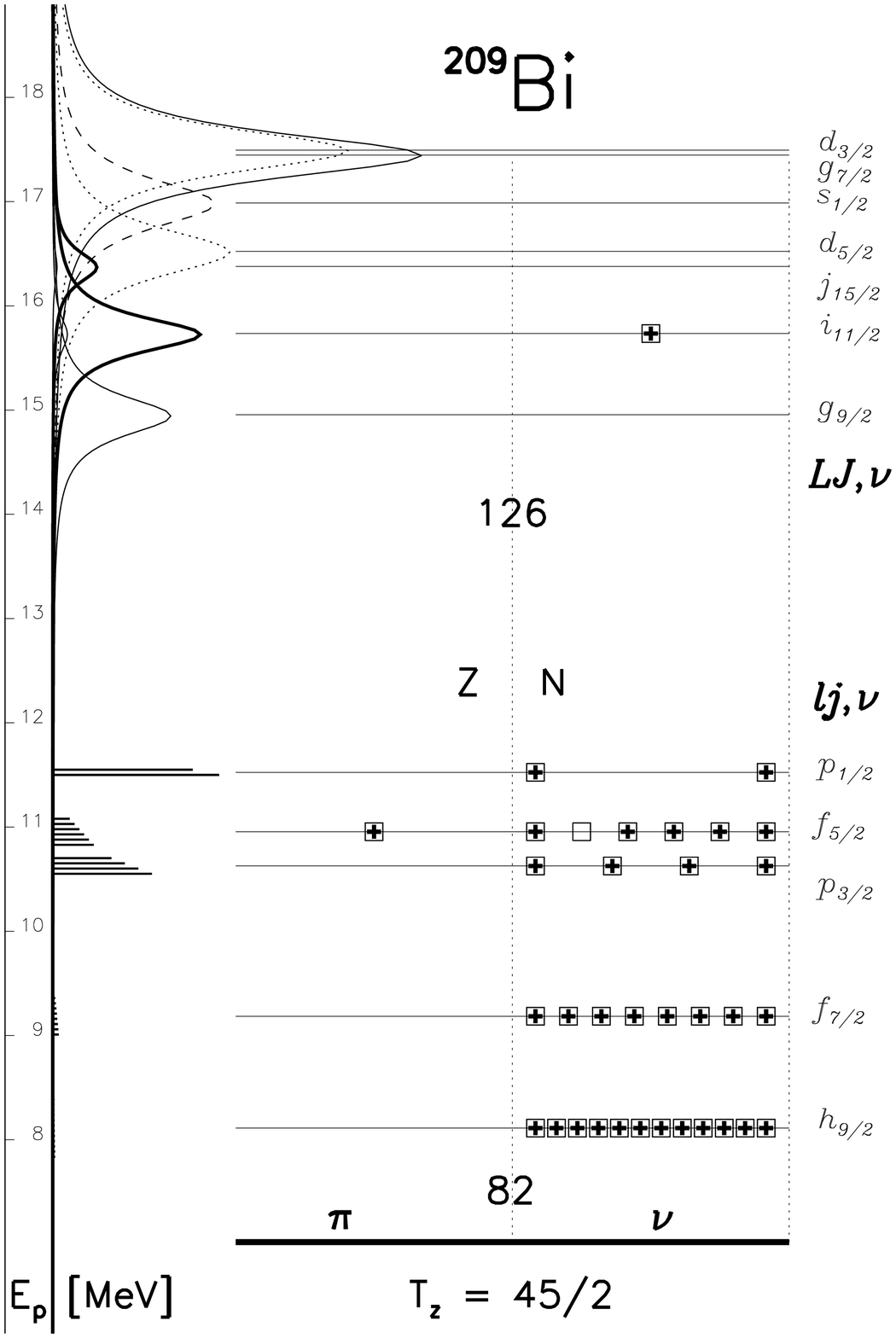}}
}
\end{figure}

In order to describe the structure of the excited states in \Pb, we
restrict the shell model wave function to the 1-particle 1-hole \cfgs,
neglecting 2-particle 2-hole and higher \cfgs.
In the restricted shell model for \Pb\  a state $|\alpha\,I>$ is
described by a superposition of \ph\ \cfgs\ built from neutrons $\nu$
and protons $\pi$ relative to the $0^{+}$ g.s. of \Pb, see 
Fig.~\ref{IAR.scenario} for the neutron particle and hole \cfgs\
$LJ,\nu$ and $lj,\nu$,
\begin{eqnarray}
\label{eq.state}
|\alpha I> = 
\sum_{LJ} \sum_{lj} c_{LJ,lj}^{\alpha\,I,\nu}
|LJ{,\nu}> \otimes|lj{,\nu}>
+ 
\nonumber\\
\sum_{LJ} \sum_{lj} c_{LJ,lj}^{\alpha\,I,\pi}
|LJ{,\pi}> \otimes|lj{,\pi}>.
\end{eqnarray}
Here
we characterize a state $|\alpha\, I>$ by its spin (always given
together with the parity), $\alpha$ denoting 
the  excitation energy $E_x$ and other quantum numbers.
The excitation energy is often given as a {\it label} by using 
\cite{Schr1997} where known and omitting fractions of keV,
{\it so an adopted value of the energy may differ by 1~keV}. 
If we restrict to this ansatz
the amplitudes  $c_{LJ,lj}^{\alpha\,I,(\nu,\pi)}$ represent a unitary
transformation of the  shell model \ph\ \cfgs\ 
to the real states $|\alpha\,I>$.

We introduce the short-hand writing 
$|LJ,\nu>$ for the neutron particle in the 6th shell 
with angular momentum $L$ and spin $J$ and similarly 
$|lj,\nu>$ for the neutron hole     in the 5th shell,
$|LJ,\pi>$ for the proton particle in the 5th shell,
$|lj,\pi>$ for the proton hole     in the 4th shell.
We often omit the label $\nu$ and simply write e.g. $\dFhlb$ 
since we 
will  essentially only discuss neutron \ph\ \cfgs\ in this paper.
From the context e.g. the meaning of the neutron particle
$|LJ,\nu>=6\,\dFhlb$ can be distinguished from the proton hole
$|lj,\pi>=4\,\dFhlb$.

In the schematic shell model (SSM) the  residual interaction is taken
to be zero. 
The  splitting of the multiplets in the full shell model depends on
the strength of the  
diagonal and nondiagonal matrix elements of the residual interaction
in \Pb\ and of the relative separation of the  undisturbed 
\cfgs\ in the SSM;
matrix elements in the order of 
magnitude of some tens of keV are expected \cite{AB1973}.

A rather pure structure will show up only in case of an isolated
multiplet.
Actually, however, the lowest 20 states in \Pb\ ($E_x<$4.5\,MeV) are heavily
mixed since here the \hNhlb\sOhlb\ proton and the \gNhlb\fFhlb\
neutron \cfgs\ have almost the same SSM energy, and similarly
the \hNhlb\dThlb\ proton and the \gNhlb\pThlb, \iEhlb\pOhlb\ neutron
\cfgs.
An early attempt \cite{AB1973}  determined
the matrix elements of the effective residual interaction among \ph\ 
\cfgs\ in \Pb\ 
from the \cfg\ mixing in the lowest 20~states.
However, 
some spin assignments and identifications of the states below
$E_x$=4.50\,MeV were essentially settled by the later work of
\cite{Mai1983}. In 1982, an update of the fit was done by one
of us (A.~H.); the results are shown in appendix~A.  There is a
remarkable agreement with shell model calculations by \cite{Rej1999}.

In contrast, the two  multiplets built from the \iEhlb\ neutron
particle and the \fFhlb\ and the \pThlb\ neutron hole,  
predicted  at SSM energies  $E_x$=4.780 and 5.108\,MeV, respectively,
are expected to be less mixed, at least for the high spin members
($I=5,6,7,8$). 

\section{Selective reactions
}

Spectroscopic information about \ph\ \cfgs\ has been derived in
addition to IAR-pp' from particle transfer reactions \PbS(d,~p),
\Bi\dHe, and \Bi(t,~$\alpha$) \cite{Mai1983,Schr1997,Valn2001} and
from transitions due to the electromagnetic \cite{Rad1996,Rej1999} or
the weak interaction \cite{NDS1986}.  
IAR-pp' allows to identify the neutron components
$|LJ{,\nu}>\otimes|lj{,\nu}>$ of \ph\ states. The quantum number of
the selected IAR is identical to the quantum number of the neutron
particle configuration $|LJ{,\nu}>$, the angular distribution of the
inelastically detected protons 
carry the information on the coherent contribution of the excess
neutrons
$|lj{,\nu}>$.

\subsection{IAR-pp'
}

We discuss the inelastic proton scattering via isobaric analog
resonances ({\it IAR-pp'\,}) on a spin~0 target, ${|0^{+}g.s.>} 
\rightarrow IAR(LJ) \rightarrow {|\alpha\,I>} $, here
specifically \Pb(p,~p') proceeding via one of the lowest, well
isolated IAR in \Bi.

The wave function of an IAR in \Bi\ with spin $LJ$ may be represented
by
\begin{eqnarray}
\label{eq.IAR.doorway}
|\Psi_{LJ}^{IAR}(\Bi)> =
\nonumber\\
{\frac{1}{\sqrt{2T_0+1}}}
T_{-}
|LJ,\nu>\otimes|\Pb(0^{+}\, g.s.)>
\end{eqnarray}
where $T_0=(N-Z)/2$ is the isospin of the g.s. of \Pb.
The isospin lowering operator $T_{-}$ acts on all excess neutrons,
hence we have
\begin{eqnarray}
\label{eq.IAR.sum}
|\Psi_{LJ}^{IAR}(\Bi)> =
\nonumber\\
{\frac{1}{\sqrt{2T_0+1}}}
|LJ,\pi>
\otimes|\Pb(0^{+}\, g.s.)> +
\nonumber\\
\sum_{lj} 
\sqrt{\frac{2j +1}{2T_0+1}}
(|lj^{+1},\pi> \otimes |lj^{-1},\nu >)_{0^{+}}
\nonumber\\
\otimes |LJ,\nu> \otimes
|\Pb(0^{+}\, g.s.)>
.
\end{eqnarray}
Evidently the outgoing proton either leaves \Pb\ in its g.s. 
(elastic scattering) or creates a {\it neutron} \ph\ \cfg
$|LJ,\nu>\otimes|lj,\nu>$ as shown in the sketch
Fig.~\ref{IAR.scenario} for one specific example.

\subsection{\AngDs\ of \Pb(p,~p')
}
The IAR are described as Breit Wigner like resonance terms, their
partial decay widths depend on the mixing coefficients
$c_{LJ,lj}^{\alpha\,I}$ and on \peneTra\ effects.
The resonance scattering is nicely described in
the book of Bohr\&Mottelson \cite{BM1969} in a general manner.

The differential cross section of the \Pb(p,~p') reaction  on top of
an isolated IAR 
($E_p=E^{res}_{LJ}$) proceeding to a state with  neutron \ph\
\cfgs\ $|LJ>\otimes|lj,\nu>$ is described  \cite{Heu1969} by
\begin{eqnarray}
\label{eq.diff.c.s}
{\frac{{d\sigma_{LJ}^{\alpha\,I}}} {d\Omega}}(\Theta)
 = 
{\frac{\hbar^2 }{4\mu_0 }} 
{\frac{(2I+1)}{(2J+1)}}
{\frac{\Gamma^{s.p.}_{LJ}}{{E^{res}_{LJ}(\Gamma^{tot}_{LJ})^2}}}
\,\times
\nonumber\\
\sum_{lj}\sum_{l'j'}
a^{IK}_{LJ,lj,l'j'}
P_K(cos(\Theta))
{c_{LJ,lj}^{\alpha\,I}\sqrt{\Gamma^{s.p.}_{lj}}}
\nonumber\\
cos(\xi^{s.p.}_{lj}-
    \xi^{s.p.}_{l'j'})
{c_{LJ,l'j'}^{\alpha\,I}\sqrt{\Gamma^{s.p.}_{l'j'}}}
\end{eqnarray}
where $\xi^{s.p.}_{lj}$ are phases derived from theory
\cite{1971CL02} and 
$\mu_0=m(p)m(\Pb)/(m(p)+m(\Pb))$ is the reduced mass.
The factors $a^{IK}_{LJ,lj,l'j'}$ arise from the
recoupling of the angular momenta $L,l$ and spins $J,j$ to $I,K$
\begin{eqnarray}
a^{IK}_{LJ,lj,l'j'}
=
(-)^{(I+2J)}
W(jJj'J,IK)
\,\times
\nonumber\\
 \bar Z(LJLJ,{\frac{1}{2}} K)
 \bar Z(ljl'j',{\frac{1}{2}} K),
\end{eqnarray}
where 
$K\le min(2L,2J,max(2l),max(2j))$ is even; the recoupling \coefs\ 
$W, \bar Z$ are defined by \cite{1952BB,Edm1957}, 
see appendix~B. 

The component with $K=0$ represents the mean cross section
$\sigma^{\alpha\,I}_{LJ}$;
for a  state $|\alpha\,I>$ 
it is just the sum
of the \cfg\ strength
$|{c_{LJ,lj}^{\alpha\,I}}|^2$
weighted by the s.p. widths, 
\begin{eqnarray}
\label{eq.avg.c.s}
\sigma^{\alpha\,I}_{LJ} = 
{\frac{\hbar^2}{4\mu_0 E^{res}_{LJ}}} {\rm  } 
{\frac{(2I+1)}{(2J+1)}}
{\frac{\Gamma^{s.p.}_{LJ}}{{(\Gamma^{tot}_{LJ})^2}}}
\sum_{lj}
|{c_{LJ,lj}^{\alpha\,I}\sqrt{\Gamma^{s.p.}_{lj}}}|^2.
\end{eqnarray}
For a multiplet of states $|\alpha\,I>$ 
with spins $I=J-j,\cdots,J+j$ consisting 
mainly of one \cfg\
$|LJ>\otimes|lj>$, the angle averaged (mean) cross sections $\sigma^{\alpha\,I}_{LJ}$
on top of a specific IAR $LJ$ 
should be simply related to the spin factor $2I+1$, 
neglecting contributions from other IAR.

In general several \cfgs\ are to be considered; the
formula describing the \angD\ of the IAR-pp' reaction
(Eq.~\ref{eq.diff.c.s})
comprises a sum of  products for coherent amplitudes
$c_{LJ,lj}^{\alpha\,I}$. 
Hence the relative phases of the amplitudes can be determined.

Each pure neutron \ph\ \cfg\ $|(LJ{,\nu}>\otimes|lj{,\nu}>)_I>$ has a
characteristic \angD\ $\sum_K a_K P_K(\Theta)$ of even Legendre
polynomials (appendix~B).  Small admixtures of other neutron \ph\
\cfgs\  sometimes change the values $a_K$ considerably, however.

The highest spin of each \cfg\ $|LJ>\otimes|lj>$ produce a deep
minimum of the \angD\ at $90^\circ$, which is the more pronounced the
higher the angular momenta $LJ,lj$ are.  Similarly, for the lowest
spin a deep minimum of the \angD\ at $90^\circ$ is found.
This gives the chance to assign spins in certain cases rather firmly.

Unfortunately we could not measure at scattering angles beyond
$115^\circ$ for technical reasons.  IAR-pp' \angDs\ should be
symmetric around $90^\circ$ in the absence of direct-(p,~p') contributions.
Hence an \angD\ rising towards forward angles sometimes is difficult
to interprete.

In case a group of states represents a rather complete subset of \ph\
\cfgs, the \coefs\ $c_{LJ,lj}^{\alpha\,I,(\pi,\nu)}$ of the unitary
transformation matrix may be determined from the analysis of IAR-pp'
together with the observation of the \ONrule\ and the sum-rule
relations.
Often there are less free parameters to be fitted than the IAR-pp'
data  provides.
So in principle, amplitudes of proton \ph\ \cfgs\ can be determined
\cite{AB1973}.

Crucial for such an analysis is the correct identification of all
relevant states and firm spin and parity assignments.

\begin{table}[htb]
\caption[IAR parameters
]
{\label{ratio.spWid}%
Parameters for IAR in \Bi. For some IAR $LJ$ and some outgoing waves
$lj$ new values are derived (rightmost column), see appendix~C.
The energy dependence of the \peneTra\ for the escape widths
$\Gamma_{LJ}^{s.p.}$ can be globally approximated by
Eq.~\ref{eq.peneTra} in the region 10\,MeV$<E_{p'}<12$\,MeV.
}
\begin{tabular}{|c ccc c| c |}
\hline
$LJ$ &$E_{LJ}^{res}$ & $\Gamma_{LJ}^{tot}$& $\Gamma_{LJ}^{s.p.}$&
$R_{LJ}$ &$\Gamma_{LJ}^{s.p.}$ \\ 
     & MeV      & keV            & keV             &      &keV\\
&\cite{1968WH02}&\cite{1968WH02}&\cite{1968WH02}   &\cite{1968WH02} &   \\
\hline
\gNhlb & 14.918$\pm$.006 & 253$\pm$10 & 20$\pm$ 1 &  8    &     \\
\iEhlb & 15.716$\pm$.010 & 224$\pm$20 & ~2$\pm$0.8&  1    & ~2.2$\pm$0.3~~ \\
\jFhlb & 16.336$\pm$.015 & 201$\pm$25 &           &0.4${}^a$& ~0.7$\pm$0.3 \footnote{
from a preliminary analysis of 
the 4610, 4860, 4867  states with
spins $8^{+}, 8^{+}, 7^{+}$} \\ 
\dFhlb & 16.496$\pm$.008 & 308$\pm$~8 & 45$\pm$ 5 & 12    &     \\
\sOhlb & 16.965$\pm$.014 & 319$\pm$15 & 45$\pm$ 8 & 11    &     \\
\dThlb & 17.430$\pm$.010 & 288$\pm$20 & 35$\pm$10 & (20)
\footnote{doublet IAR, definition of $R_{LJ}$ valid for isolated IAR
only}                         &     \\
\gShlb & 17.476$\pm$.010 & 279$\pm$20 & 45$\pm$10 & (20) ${}^b$ &\\
\hline
 $lj$ &$E_{lj}^{p'}$ & &$\Gamma_{lj}^{s.p.}$ &&$\Gamma_{lj}^{s.p.}$\\
      & MeV      & & keV                     &&keV \\ 
 &\cite{1969RI10} \footnote{
$E_{lj}^{p'} = E_{LJ}^{res}-E_{LJ,lj}^{SSM}$
corresponds to the SSM energy of the \ph\ \cfg\ $|LJ>\otimes|lj>$,
see Fig.~\ref{IAR.scenario}.
              } & &\cite{1969RI10}          &&    \\
\hline
 \pOhlb & 11.49 & &28.6$\pm$3& &28.6   \footnote{
This value was not adjusted since the systematic errors of the
absolute cross section are about 10-20\%.
They can be reduced by a more complete evaluation
of our IAR-pp' data \cite{AHwwwHOME}.
}~~~ \\
 \fFhlb & 10.92 & &~4.2$\pm$0.4& & ~~~5.2$\pm$0.4~~  \\
 \pThlb & 10.59 & &15.8$\pm$1.5& &~~14.6$\pm$0.5~~  \\
 \fShlb & ~9.15 & &~0.6~~~~~   & &~~~0.55$\pm$0.1  \footnote{
from a preliminary analysis of the 5935 state
identified to contain most of the \gNhlb\fShlb\ $8^{-}$ \cfg}\\
\hline
\end{tabular}
\end{table}

\subsection{Energy dependence of the s.p. widths
}
The s.p. widths strongly depend on the angular momentum $L$
of the IAR since the outgoing particle has to penetrate the Coulomb
barrier. Fig.~\ref{IAR.scenario} gives an impression about the
relative values of the \peneTra.  The \iEhlb\ IAR with $l=4$ has the
weakest \peneTra\ of all positive parity IAR we measured.

We define a \peneTra\ ratio
\begin{eqnarray}
\label{eq.ratio}
R_{LJ}= 
{\frac
{\Gamma^{s.p}_{LJ}/(\Gamma^{tot}_{LJ})^2}
{\Gamma^{s.p}_{\iEhlb}/(\Gamma^{tot}_{\iEhlb})^2}
};
\end{eqnarray}
it
compares the cross section on  some IAR $LJ$ to that
on the \iEhlb\ IAR. In fact it essentially takes care of the different
\peneTra\ of the particle populating each IAR.
Using the data and analysis of   \cite{1968WH02,1969RI10} 
we derive  values 
$R_{LJ}$=8, 12,~11 for $LJ$=\gNhlb, \dFhlb, \sOhlb\ IAR, respectively,
see Tab.~\ref{ratio.spWid}.
For the doublet \dThlb+\gShlb\ IAR we assume a factor 20, but we note 
that the given equations are valid for isolated IAR
\cite{Heu1969a,Latz1979}  only.

\subsection{Overlapping IAR
}

Eqs. \ref{eq.diff.c.s},~\ref{eq.avg.c.s} are valid for isolated IAR
only. The lowest IAR in \Bi\ are well isolated, but
the \iEhlb\ IAR is rather weak as can be seen  
in Fig.~\ref{IAR.scenario} where it is enhanced by a 
factor~10 in order to make it visible at all.
Hence  the tails from neighbouring \gNhlb\ and \dFhlb\ IAR may
interfere with the \iEhlb\ IAR.
(The \peneTra\ ratio is $R_{\gNhlb}, R_{\dFhlb}$=8,~12.)
Following the formula for \excFs\ given by \cite{1968WH02}, 
the \gNhlb, \dFhlb\ IAR have decayed by a factor 40,~25,
respectively, from  
the top of the IAR ($E_p$=14.920, 16.945\,MeV, respectively) to
$E_p$=15.720\,MeV,  the resonance energy of the \iEhlb\ IAR.
Since in IAR-pp' the amplitudes are relevant,
the population on top of the \iEhlb\ IAR by the
neighbouring IAR can be still considerable. However, the influence of
the \gNhlb\ IAR may be neglected since the relative amplitude is only
of the order of 15\%, while for the \dFhlb\ IAR the relative amplitude
is still about~50\%.

Yet for the \iEhlb\fFhlb, \iEhlb\pThlb\ multiplets being considered,
this problem does not apply since there must be an allowed entrance
channel. For the higher spins only the \cfgs\ \dFhlb\hNhlb,
\dFhlb\hEhlb\ may contribute, but the \peneTra\  of the outgoing
$l=5$ particle is 10 and 50 times lower than for \fFhlb\ and
\pThlb, respectively. In addition any contribution of these \cfgs\ is
expected to be small.  Only for a $3^{-}$ state the  entrance channel 
\dFhlb\pOhlb\ may contribute eventually.

We conclude that IAR-pp' is a method able to detect  and  analyze even
weakly excited neutron \ph\ states.

\begin{table}[htb]
\caption[
Q3D Parameters IAR-pp' and \PbS(d,~p)
]
{\label{Q3D.params.pp}%
Parameters for the \Pb(p,~p') \expt.
Targets enriched in \Pb\ to 99.85\% were used. 
The thickness of the targets  T2, T3, T4  were
98, 245, 353\,$\mu\,g/cm^2$;
the thickness of target T1  was determined as 104\,$\mu\,g/cm^2$ 
by comparison to other targets.
}
\begin{tabular}{|c|llcc| l|}
\hline
IAR &$E_p [MeV]$ &$E_x [MeV]$ &$\Theta$ &targets & \# runs  \\
\hline
\gNhlb &14.920 &3.85 - 6.2& $48^\circ$ - $115^\circ$ & T1 - T4 & 57 
\footnote{
2\,runs at $\Theta=54^\circ,90^\circ$ covering  $E_x$=2.1-3.85\,MeV}
\footnote{
 1\,run at $\Theta=58^\circ$ covering  $E_x$=6.2-6.65\,MeV}
\\
\iEhlb &15.720 &4.05 - 5.85& $20^\circ$ - $115^\circ$ & T1 - T3 & 44 
\footnote{
3\,runs at $\Theta=105^\circ,115^\circ$ covering
$E_x$=3.85 - 4.05\,MeV,}
\footnote{
 1\,run at $\Theta=105^\circ$ covering  $E_x$=5.85 - 6.18\,MeV }
\\
\jFhlb &16.355
\footnote{in addition 16.290, 16.380, 16.290}
 &4.55 - 6.0& $66^\circ$ - $115^\circ$ & T2 - T3& 22 \\
\dFhlb &16.495 &3.73 - 6.9& $36^\circ$ - $115^\circ$ & T1 - T4 & 39 
\footnote{
1\,run at $\Theta=48^\circ$ covering  $E_x$=3.65 - 3.73\,MeV}
\footnote{
 3\,runs at $\Theta=48^\circ,84^\circ$ covering  $E_x$=6.9 - 7.4\,MeV}
\\
\sOhlb &16.960 &5.00 - 6.9& $48^\circ$ - $115^\circ$ & T2 - T4 & 12
\footnote{
  1\,run  at $\Theta=84^\circ$ for $E_x$=3.65 - 5.0\,MeV}
\footnote{
 1\,run  at $\Theta=115^\circ$ for $E_x$=6.9 - 7.2\,MeV}
\\
\dThlb+\gShlb&17.480
\footnote{in addition 17.590, 17.610, 17.720}
 &5.54 - 6.8& $84^\circ,115^\circ$ & T2 - T3 & 12 
\footnote{
2\,runs at $\Theta=84^\circ$ covering $E_x$=4.7 - 5.54\,MeV}
\footnote{
 2\,runs at $\Theta=84^\circ$ covering  $E_x$=6.8 - 7.2\,MeV}
\\
\hline
\end{tabular}
\end{table}

\begin{table}[htb]
\caption[
Q3D Parameters  \PbS(d,~p)
]
{\label{Q3D.params.dp}%
Parameters for the \Pb(d,~p) \expt\ with the Q3D facility. 
The deuteron energy was $E_d$=22.000\,MeV as for \cite{Valn2001}. 
A target enriched in \PbS\ to 99.86$\pm$.04\% was used. 
The  slits perpendicular to the scattering angle were kept open,
$\Delta\Phi=\pm3^\circ$.
}
\begin{tabular}{|ccc|c|}
\hline
$E_x$ &scattering     & slit         & \# runs  \\
$[MeV]$ & angle $\Theta$& opening $\Delta\Theta$&\\
\hline
3.5  - 5.2  &$20^\circ$  & $\pm0.9^\circ$& 1\\
3.1  - 7.9  &$20^\circ$  & $\pm1.5^\circ$& 3\\
\hline
3.1  - 5.5  &$25^\circ$  & $\pm0.9^\circ$& 3\\
3.1  - 7.9  &$25^\circ$  & $\pm1.5^\circ$& 3\\
\hline
3.1  - 5.1  &$30^\circ$  & $\pm0.6^\circ$& 1\\
3.1  - 5.2  &$30^\circ$  & $\pm0.9^\circ$& 1\\
5.7  - 8.0  &$30^\circ$  & $\pm0.9^\circ$& 2\\
3.1  - 8.0  &$30^\circ$  & $\pm1.5^\circ$& 6\\
\hline
\end{tabular}
\end{table}


\section{Experiments
}

We performed \expt s on \Pb(p,~p') and \PbS(d,~p).
The high Q-value of the reaction \Bi\dHe\ prohibited any reasonable 
\expt\ with the Q3D facility due to the restricted  energy range of
the accelerator.

The data are evaluated by help of the computer code GASPAN
\cite{Rie2005}. 
It allows the deconvolution of spectra into a set of peaks with
gaussian shape of individual widths and exponential tails on a
background with polynomial shape.
The energy calibration takes care of the quadratic dependence on the
channel due to the effect of the magnetic field.

Here we report on results leading to the detection  of the main
components of the 
\iEhlb\fFhlb\ and the \iEhlb\pThlb\ multiplets in \Pb. 
Other data are being evaluated; the raw data (together with excerpts
from the runbook) can be accessed \cite{AHwwwHOME}.

A preliminary analysis is in agreement with data from 
ref.~\cite{1967MO25,1967RI13,1968BO16,1968CR05,1968VO02,1968WH02,
Zai1968,1969RI10,1970KU13} obtained in the 1960s.  Of course,
because of the much higher resolution many levels are resolved to be
doublets. An important difference is the better energy calibration,
their energies deviate linearly by about 5-10\,keV for the range
4-6\,MeV; mostly the energies are about 0.2\% too low.

\subsection{\Pb(p,~p') \expt\ with the Q3D facility
}

The \Pb(p,~p') \expt\ was performed with a proton beam from the
M\"unchen HVEC MP Tandem accelerator using the Q3D magnetic
spectrograph.
The bright Stern-Gerlach polarized ion source was used with
unpolarized hydrogen \cite{LMUrep2000p70,RalfsQ2005}.  
At beam intensities of about 900\,nA, the target was wobbled with a
frequency of 2~sec to avoid damage of the lead target.
The proton energies were chosen according to \cite{1968WH02} to match
the top of the seven lowest IAR in \Bi, namely 
the \gNhlb, \iEhlb, \jFhlb, \dFhlb, \sOhlb\ IAR and the doublet-IAR
\gShlb+\dThlb;  some more energies
slightly off-resonance were chosen, see Tab.~\ref{Q3D.params.pp}. 
The analyzed particles were detected in an ASIC supported cathode strip
detector \cite{LMUrep2000p71,Wirth1999}.
At an active length of 890~mm it produces spectra where the position
of a line is determined to better than 0.1~mm without systematic
errors.  With a few exceptions the slits of the magnetic spectrograph
were kept open, $\Delta\Theta=\pm3^\circ$, $\Delta\phi=\pm3^\circ$.

\subsection{Experiments on \PbS(d,~p)
}

A weak excitation by \PbS(d,~p) may help to decide some spin and \cfg\
assignments. 
Therefore, we measured the reaction \PbS(d,~p) with the goal to
detect as low spectroscopic factors (S.F.) as possible. 
We performed two measurements, one with the 
(gone) Buechner  spectrograph at Heidelberg
at large backward angle in order to eliminate any contamination
from light nuclei in the spectrum,
another \expt\ with the Q3D facility at M\"unchen were the deuteron
energy was chosen to match  \cite{Valn2001}. 

{\it (a) Buechner spectrograph.}
In 1969, using the Heidelberg Tandem van de Graaff accelerator, two of
us (A.~H., P. von~B.) did
a deep exposure of \PbS(d,~p) with the Buechner magnetic spectrograph
gathering 6\,mCb of the deuteron beam in more than 30 hours.
The scattering angle was chosen as $\Theta=130^\circ$.
The target was enriched to 92\%.
A short exposure  was done to position the line from the 3.708\,MeV
$5^{-}$ state properly.
The energy range was 3.65\,MeV$<E_x<5.15$\,MeV.
This data was crucial for the fit shown in \cite{AB1973} and now is
still useful albeit the re\-so\-lu\-tion of only 12\,keV.
It has been reevaluated by use of the GASPAN code \cite{Rie2005}.

{\it (b) Q3D spectrograph.}
The study of the reaction \PbS(d,~p) was performed with a deuteron beam
from the M\"unchen HVEC MP Tandem accelerator.
The high performance of the Q3D facility allowed to take 18~spectra
with superior resolution during 30~hours with beam intensities of
about 600\,nA.  Tab.~\ref{Q3D.params.dp} shows the parameters relevant
to the data taking.
                               

In order to detect even minor contaminations (e.g. from
\Nax,\ClF,\ClS) we measured at scattering angles
$\Theta=20^\circ,25^\circ,30^\circ$ with different slit openings, see
Tab.\ref{Q3D.params.dp}. 
We achieved  a peak-to-valley ratio of better than $1:10^{-4}$ which
allows  the detection of S.F. as low as a few $10^{-3}$ in favorable
cases. 
By this means,
the amount of the impurity isotopes \PbX, \Pb\ could be measured as
0.028$\pm$.003\%,
0.11$\pm$.03\%, respectively.

The 5292, 4610\,keV levels in \Pb\
(Tab.~\ref{allStates.other.1},~\ref{allStates.other.2}) are known to
be populated by a $l=0,\,l=5$ transfer, respectively.  The measurement
at three scattering angles allows to discriminate the transfer of a
$l=0,\,l=5$ neutron by virtue of a steeply rising slope
for the \angD.
%
This gives a chance to determine the $l$-value for some levels. 
Other $l$-values have about equal cross sections for
$\Theta=20^\circ,25^\circ,30^\circ$. 

\begin{figure}[htb]
\caption[
Spectra for $E_x$=4.6-5.0\,MeV
]
{\label{Pb_pp_46.50}%
Spectra of \Pb(p,~p') for $E_x$=4.6-5.0\,MeV 
taken at $\Theta=58^\circ,72^\circ,54^\circ$ on the
\gNhlb, \iEhlb, \dFhlb,
with targets T3, T2, T2 (see caption of Tab.~\ref{Q3D.params.pp}),
 respectively.
Six levels resonate at $E_p$=15.72\,MeV on top of the \iEhlb\ IAR
(black fill out); the doublet at 4709, 4711\,keV is
resolved by the computer code GASPAN only.
The energies of the \iEhlb\fFhlb\ multiplet are given in the middle
panel  and shown by bars above and below; 
in the lower panel the spins are given, too.
The counting interval is proportional to $\sqrt{E_x}$ and
one step corresponds to  about 0.3\,keV. 
}
\resizebox{\hsize}{05.05cm}{
{\includegraphics[angle=00]{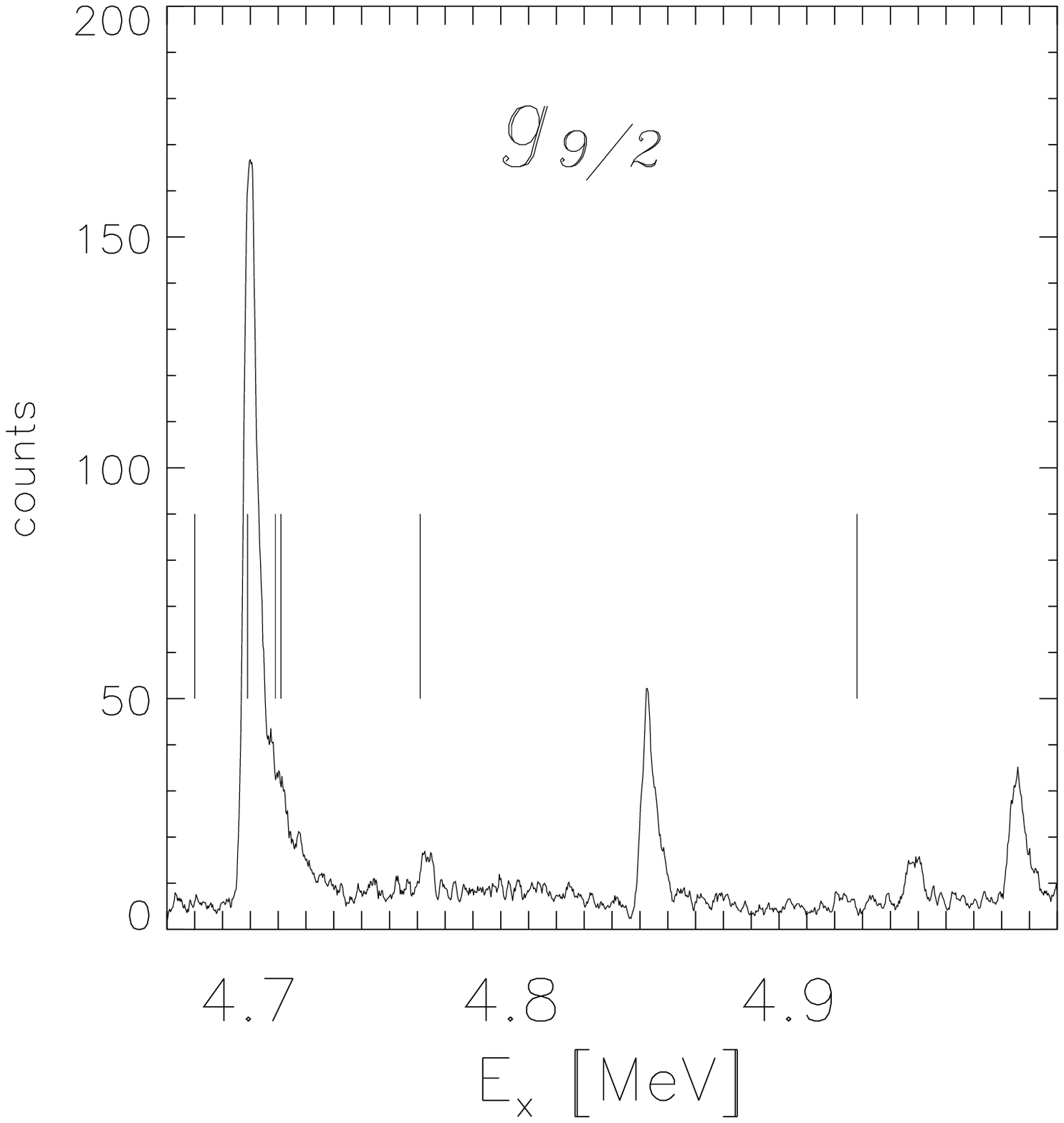}}
}
\resizebox{\hsize}{05.05cm}{
{\includegraphics[angle=00]{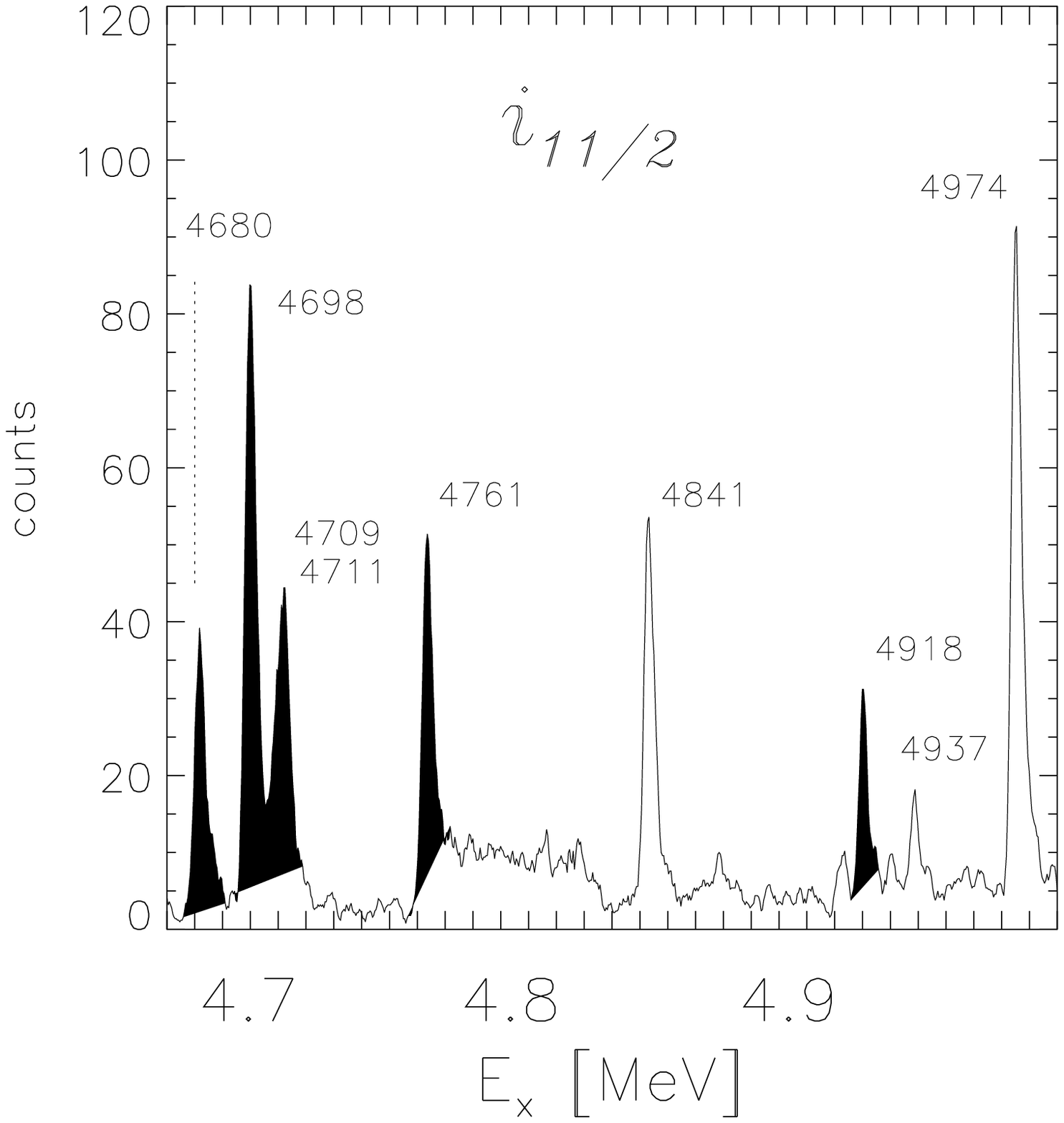}}
}
\resizebox{\hsize}{05.05cm}{
{\includegraphics[angle=00]{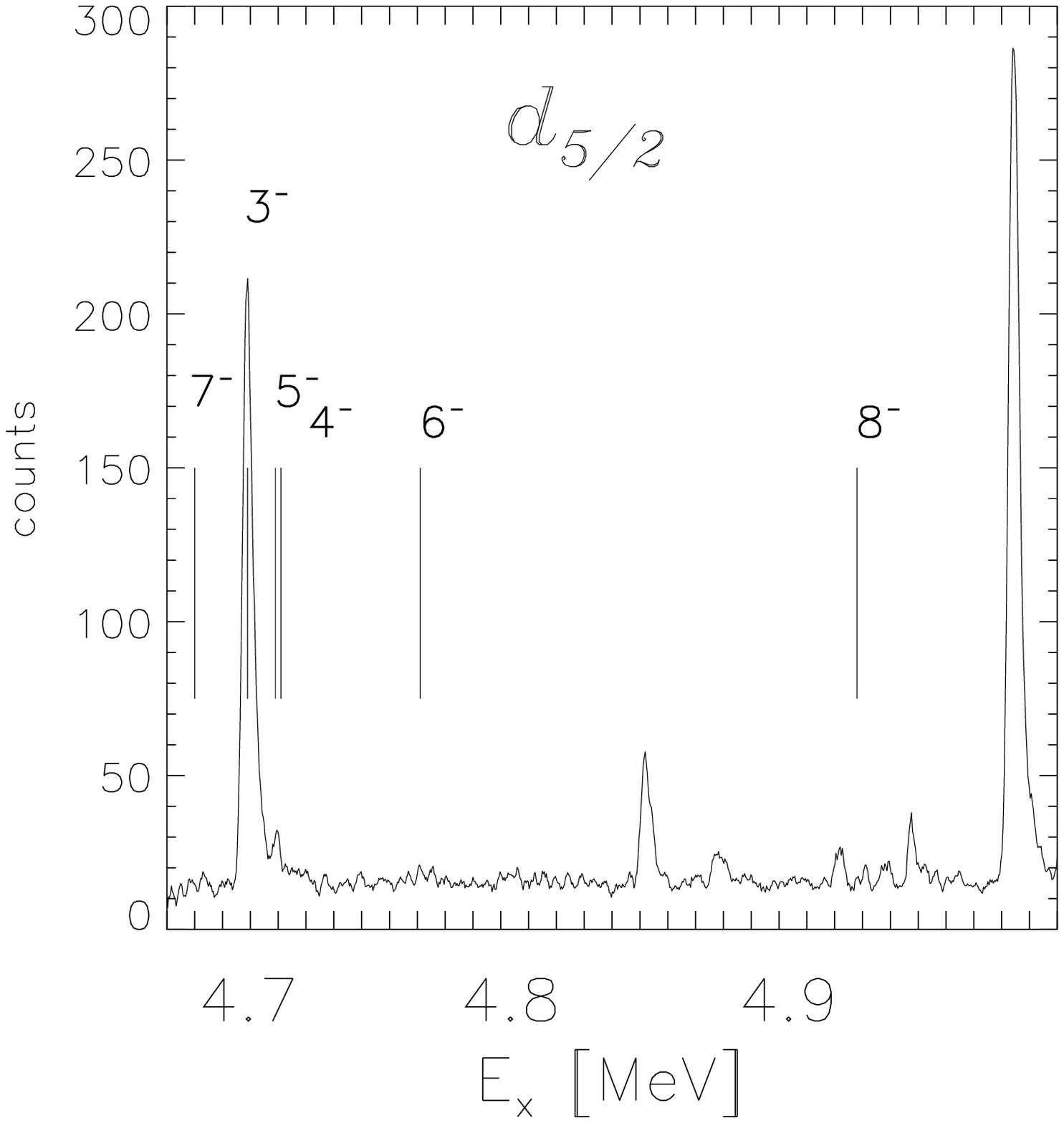}}
}
\end{figure}


\begin{figure}[htb]
\caption[
Spectra for $E_x$=5.0-5.3\,MeV
]
{\label{Pb_pp_50.53}%
Spectra of \Pb(p,~p') for the \iEhlb\pThlb\ multiplet in the region
$E_x$=5.0-5.3\,MeV.   The energies of the multiplet are given in the
middle panel, the spins in the lower panel. For other details refer to
Fig.~\ref{Pb_pp_46.50} and the text.
}
\resizebox{\hsize}{05.05cm}{
{\includegraphics[angle=00]{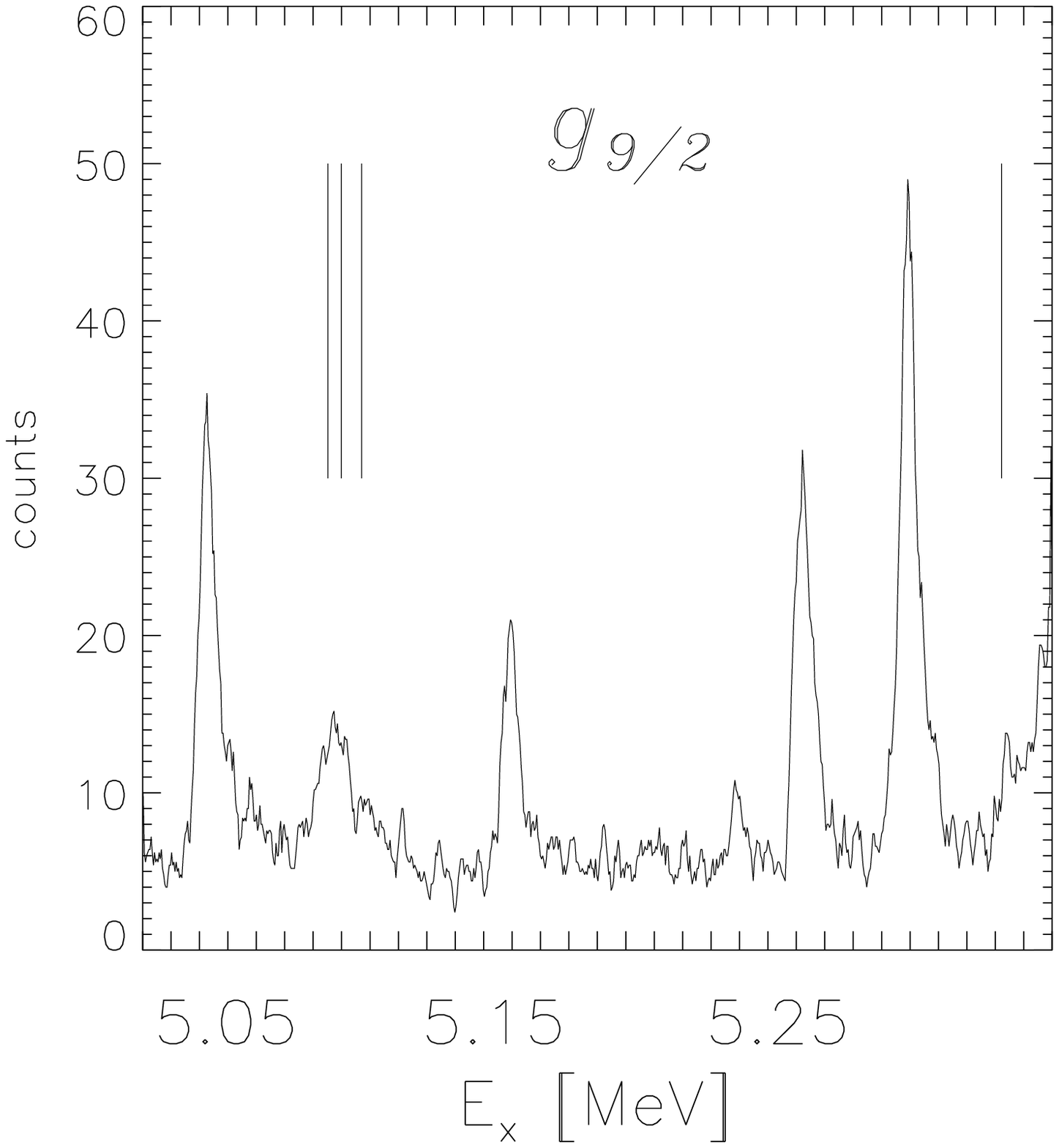}}
}
\resizebox{\hsize}{05.05cm}{
{\includegraphics[angle=00]{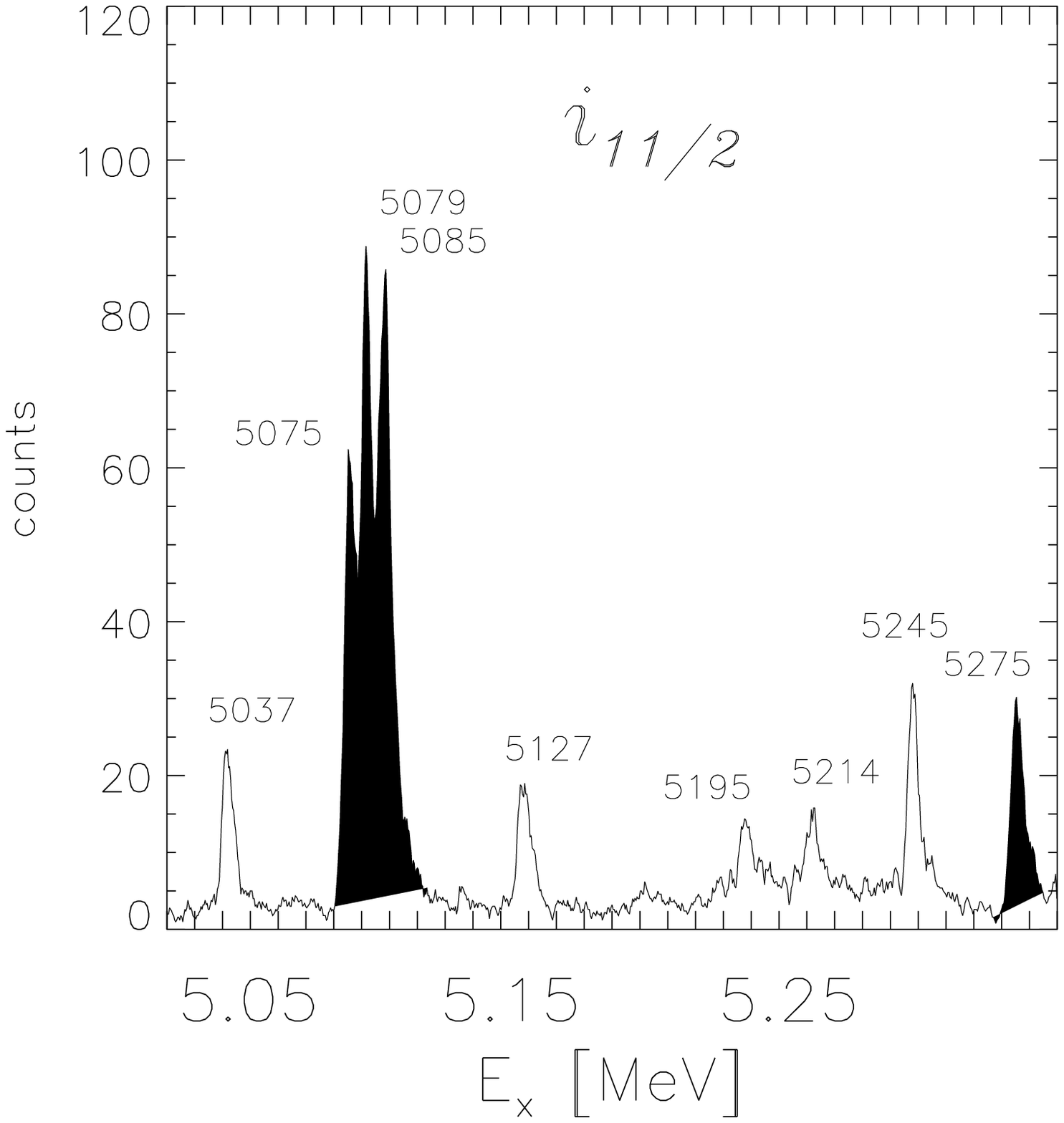}}
}
\resizebox{\hsize}{05.05cm}{
{\includegraphics[angle=00]{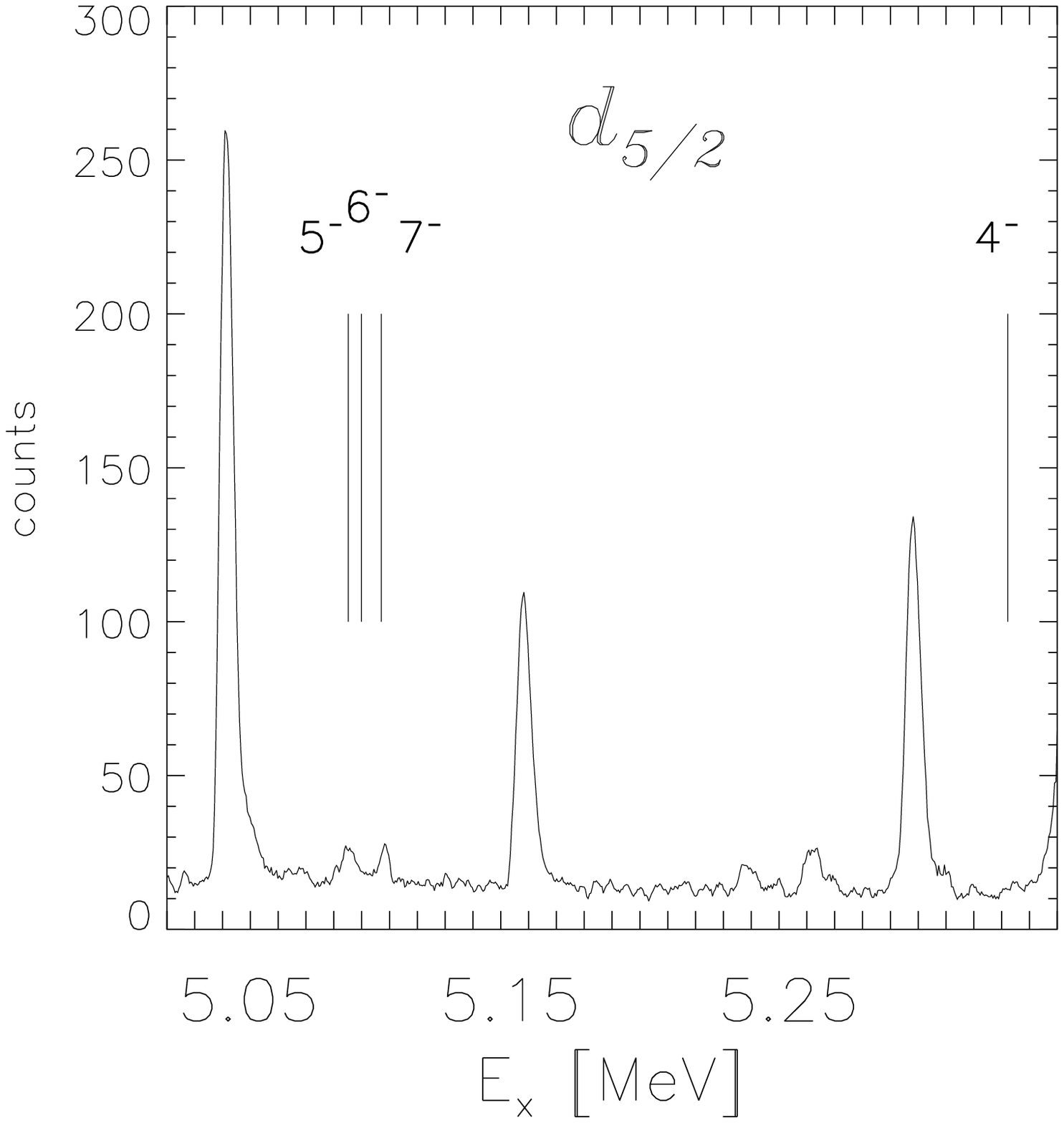}}
}
\end{figure}


\begin{table}[htb]
\caption[Energies for other states than \iEhlb\fFhlb,\pThlb -- part.1
]
{\label{allStates.other.1}%
Energies for levels excited by \Pb(p,~p')
but {\it not} bearing the main strength of the 
\cfgs\ \iEhlb\fFhlb, \iEhlb\pThlb.
Not all of them are also detected by the \PbS(d,~p) \expt.
The energies from \Pb(p,~p') are not yet finally
evaluated, therefore they are just given as a label with a
precision of around 1\,keV.
The dominant excitations by specific IAR are shown;
for strong excitations the energy label is printed {\bf boldface}, 
for weak excitations an important IAR is given in parentheses.
Energies and mean cross sections $\sigma^{\alpha I}_{LJ}$ 
($\Theta=20^\circ,25^\circ,30^\circ$) derived from \PbS(d,~p) are shown.
Spins from \cite{Schr1997} and energies from 
\cite{Valn2001,Schr1997,Rad1996}
 are given for comparison. 
}
\begin{tabular}{|cc cc|rrr c|}
\hline
$E_x$ &main &$E_x$&$\sigma(25^\circ)$ &$E_x$          &$E_x$          &$E_x$ &spin\\
      &IAR  &keV  &$\mu b/sr$         &keV            &keV            &keV   &\\
(p,~p')&     &(d,~p)& (d,~p)             &\cite{Schr1997}&\cite{Valn2001}&\cite{Rad1996}&\cite{Schr1997}\\
\hline
{\bf 5292 } &  \sOhlb & 5292.1&637  & 5292.000& 5292.6& 5292.7&$   1^{-}$ \\
           &         &    0.2&     &    0.200&    1.5&    0.1&           \\
     5280  &  \sOhlb & 5280.3&210  & 5280.322& 5281.3& 5280.3&$   0^{-}$ \\
           &         &    0.2&     &    0.080&    1.5&    0.1&           \\
{\bf 5245 } &  \dFhlb & 5245.3&713  & 5245.280& 5245.6& 5245.4&$   3^{-}$ \\
           &         &    0.2&     &    0.060&    1.5&    0.1&           \\
     5239  & (\iEhlb)& 5239.5& 10  & 5239.350& 5240.8& --~~~~&$   0^{+}$ \\
           &         &    0.7&     &    0.360&    1.5&       &           \\
     5214  &  \dFhlb & 5214.0& 45  & 5213.000& 5215.6& --~~~~&$   6^{+}$ \\
           &         &    0.3&     &    0.200&    1.5&       &           \\
     5195  &  \jFhlb & 5195.0&17   & 5195.340& 5194.3& --~~~~&$   7^{+}$ \\
           &         &    0.3&     &    0.140&    0.6&       &           \\
     5194  &  \jFhlb & --    &$< 5$& 5193.400& --~~~~& --~~~~&$   5^{+}$ \\
           &         &       &     &    0.150&       &       &           \\
{\bf 5127 } &  \dFhlb & 5127.4&682  & 5127.420& 5127.1& --~~~~& $2,3^{-}$ \\
           &         &    0.3&     &    0.090&    0.6&       &           \\
     5105  & (\iEhlb)& --    &$< 5$& --~~~~  & 5103.3& --~~~~&$        $ \\
           &         &       &     &         &    1.5&       &           \\
     5093  &  \jFhlb & 5093.2& 14  & 5093.110& 5094.3& --~~~~&$   8^{+}$ \\
           &         &    0.5&     &    0.200&    1.5&       &           \\
     5069  &  \dFhlb & --    &$< 5$& 5069.380& 5068.5& --~~~~&$  10^{+}$ \\
           &         &       &     &    0.130&    1.5&       &           \\
     5037  &  \dFhlb & 5037.4&1200 & 5037.520& 5037.2& 5037.0&$   2^{-}$ \\
           &         &    0.2&     &    0.050&    0.6&    0.1&           \\
     5010.5& (\jFhlb)& --    &$< 5$& 5010.550& 5010.0& --~~~~&$   9^{+}$ \\
           &         &       &     &    0.090&    0.6&       &           \\
\hline
\end{tabular}
\end{table}


\begin{table}[htb]
\caption[Energies for other states than \iEhlb\fFhlb,\pThlb -- part.2
]
{\label{allStates.other.2}%
$\cdots$ continuing Tab.~\ref{allStates.other.1} 
}
\begin{tabular}{|cc|cc|rrr|c|}
\hline
$E_x$ &main &$E_x$&$\sigma(25^\circ)$ &$E_x$          &$E_x$          &$E_x$ &spin\\
      &IAR  &keV  &$\mu b/sr$         &keV            &keV            &keV   &\\
(p,~p')&     &(d,~p)& (d,~p)             &\cite{Schr1997}&\cite{Valn2001}&\cite{Rad1996}&\cite{Schr1997}\\
\hline
     4992  &     (\jFhlb)& 4992.5&10   & --~~~~  & 4992.7& --~~~~&$        $ \\
           &             &    0.6&     &         &    0.6&       &           \\
{\bf  4974} &      \dFhlb & 4973.9&1350 & 4974.037& 4974.2& 4973.8&$   3^{-}$ \\
           &             &    0.2&     &    0.040&    0.6&    0.1&           \\
     4953  &     (\dFhlb)& --    &$< 5$& 4953.320& 4952.2& --~~~~&$   3^{-}$ \\
           &             &       &     &    0.230&    0.3&       &           \\
     4937  &      \dFhlb & 4937.4&33   & 4937.550& 4937.1& 4935.1&$   3^{-}$ \\
           &             &    0.4&     &    0.230&    0.3&    0.2&           \\
     4928  &     (\jFhlb)& --    &$< 5$& --~~~~  & 4928.1& --~~~~&$        $ \\
           &             &       &     &         &    1.5&       &           \\
     4911  &     (\dFhlb)& 4911.7&  6  & --~~~~  & 4910.6& --~~~~&$        $ \\
           &             &    0.5&     &         &    1.5&       &           \\
     4909  &     (\jFhlb)& --    &$< 5$& --~~~~  & --~~~~& --~~~~&$        $ \\
           &             &       &     &         &       &       &           \\
     4895  &      \jFhlb & --    &$< 5$& 4895.277& 4894.8& --~~~~&$  10^{+}$ \\
           &             &       &     &    0.080&    1.5&       &           \\
     4867  &      \jFhlb & 4868.1& 95  & 4867.816& 4866.9& --~~~~&$   7^{+}$ \\
           &             &    0.2&     &    0.080&    1.5&       &           \\
     4860  &      \jFhlb & 4860.8& 35  & 4860.840& 4859.8& --~~~~&$   8^{+}$ \\
           &             &    0.3&     &    0.080&    1.5&       &           \\
     4841  &      \dFhlb & 4841.7&22   & 4841.400& 4841.7& 4842.1&$   1^{-}$ \\
           &             &    0.4&     &    0.100&    0.3&    0.1&           \\
     4610  &      \jFhlb & 4610.7&66   & 4610.795& 4610.8& 4610.5&$   8^{+}$ \\
           &             &    0.3&     &    0.070&    0.5&    0.3&           \\
\hline
\end{tabular}
\end{table}


\subsection{Typical spectra for \Pb(p,~p')
}

In Fig.~\ref{Pb_pp_46.50},
    \ref{Pb_pp_50.53}
we show      some spectra for \Pb(p,~p') taken on the \iEhlb\ IAR.
For comparison spectra taken on  the \gNhlb, \dFhlb\
IAR are displayed, too. In total we measured nearly 200 spectra.
We will discuss the excitation of the levels at
$E_x$=4680, 4698, 4761, 4918, 5275\,keV and the clearly resolved
multiplet at $E_x$=5075, 5079, 5085\,keV
(black fill-out in the spectra taken on the  \iEhlb\ IAR,
bars on the other IAR).

The 4709, 4711 doublet is resolved by help of the computer code GASPAN
\cite{Rie2005}; the distance is found to be 1.9\,keV with an
average re\-so\-lu\-tion of somewhat less than 3.0\,keV FWHM for
spectra, see Fig.~\ref{Pb_pp_46.50}. The line contents could 
be measured quite well using a special option of GASPAN 
(fixed level distances) yielding usable \angDs.

Some excitations belong to well known levels
(Tab.~\ref{allStates.other.1},~\ref{allStates.other.2}).
A few weak lines are also clearly identified, among them are 
the $0^{+}$ state at 5239\,keV identified by \cite{Yates96c} to have
the 2-particle 2-hole structure  $
|2614\,{\rm keV}\, 3^{-}>\otimes
|2614\,{\rm keV}\, 3^{-}>$,
the $0^{-}$ state at 5280\,keV separated from the 
5276\,keV $8^{-}$ state by only 4\,keV,
the  4860, 4867\,keV doublet with spins $8^{+}$,~$7^{+}$ strongly
excited on the \jFhlb\ IAR.

For the shown spectra (Fig.~\ref{Pb_pp_46.50}, \ref{Pb_pp_50.53})
only a few contamination lines are present;
prominent contamination lines start at $E_x\approx\ $5.29\,MeV
for the spectra taken both on the \gNhlb\ and the \dFhlb\ IAR.
A weak contamination  line is visible in the region
4.76-4.82\,MeV on the \iEhlb\ IAR, kinematically broadened.

Most levels in discussion are excited strongest on the \iEhlb\ IAR,
the only exception is the 4698\,keV $3^{-}$ state.
On the \iEhlb\ IAR, the levels at 4680, 4761, 4918, 5079\,keV are at
least four times stronger excited than on any other~IAR.

\section{Data analysis
}
\subsection{Excitation energies from \Pb(p,~p')
}

In 
Tab.~\ref{allStates.info}
we show the excitation energies derived from our
measurements  of \Pb(p,~p').  The spectra were calibrated by using
around  40 reference energies below $E_x$=6.0\,MeV and about 
25 more reference energies up to $E_x$=7.5\,MeV mainly from
\cite{Schr1997}, but also from \cite{Rad1996,Valn2001};
see Tab.~\ref{allStates.other.1},~\ref{allStates.other.2} for the
region of interest.

We avoided the usage of reference values in cases where the
identification due to a multiplet structure was unclear or where the
cross section was low.
In addition to the quadratic dependence of the energy from the channel
in the Q3D spectra, a secondary fit by a third order parabola improved
the energy calibration considerably, see \cite{LMUrep2004}.

The excitation energies determined from the IAR-pp' measurement with
errors of about 0.5\,keV in general, 
compare well  to \cite{Schr1997,Rad1996,Valn2001} within the given
errors;
the only exception is the 5075 level with a discrepancy of about
two standard errors.

\subsection{Excitation functions  of \Pb(p,~p')
}

With a few exceptions, we did not measure \excFs, but
selected the energies of all known IAR only, see
Tab.~\ref{Q3D.params.pp}.   IAR often excite the states rather
selectively.  So we can determine excitation functions in a schematic
manner. For some levels \excFs\ were measured in the 1960s
\cite{Zai1968,1968WH02}. 
We will mention them in place.

The \angDs\ were fitted by even Legendre polynomials
\begin{eqnarray}
\label{eq.mean.c.s}
{\frac{{d\sigma_{LJ}^{\alpha\,I}}} {d\Omega}}(\Theta)
 = 
\sum_K A_K P_K(cos(\Theta))
\end{eqnarray}
Odd Legendre  polynomials  have not to be included since the 
direct-(p,~p') reaction does not contribute much in most cases.
The angle averaged (mean) cross section is derived for each IAR $LJ$
and each state $|\alpha\,I>$ as $\sigma^{\alpha\,I}_{LJ}=A_0$.
We don't quote neither the errors of $\sigma^{\alpha\,I}_{LJ}$ nor
the values $A_K$ for $K>0$ since the evaluation can be further
improved. The errors of the mean cross sections are about 5-20\%.

\begin{figure}[htb]
\caption[
Schematic \excFs\  i11 f5
]
{\label{IARschemiE.f5.p3}%
Angle averaged (mean) cross section $\sigma^{\alpha\,I}_{LJ}$ 
for states containing most of the 
\iEhlb\fFhlb\  strength ({\it upper panel}) and  
\iEhlb\pThlb\  strength ({\it lower panel}).
The value $\sigma^{\alpha\,I}_{LJ}$  for each state is shown 
relative to the maximum of all cross sections
of either multiplet; the maxima are set equal
(upper panel: 4761 $6^{-}$, lower panel: 5085 $7^{-}$, see
tab~\ref{allStates.info}). 
At the left and right side the energy labels and the spins are given.
In order to obtain partial widths (Eq.~\ref{eq.avg.c.s}),
for each IAR $LJ$ the mean cross section must be reduced by the
\peneTra\ ratio $R_{LJ}$ given at bottom.
}
%
%
\resizebox{\hsize}{08.4cm}{
{\includegraphics[angle=00]{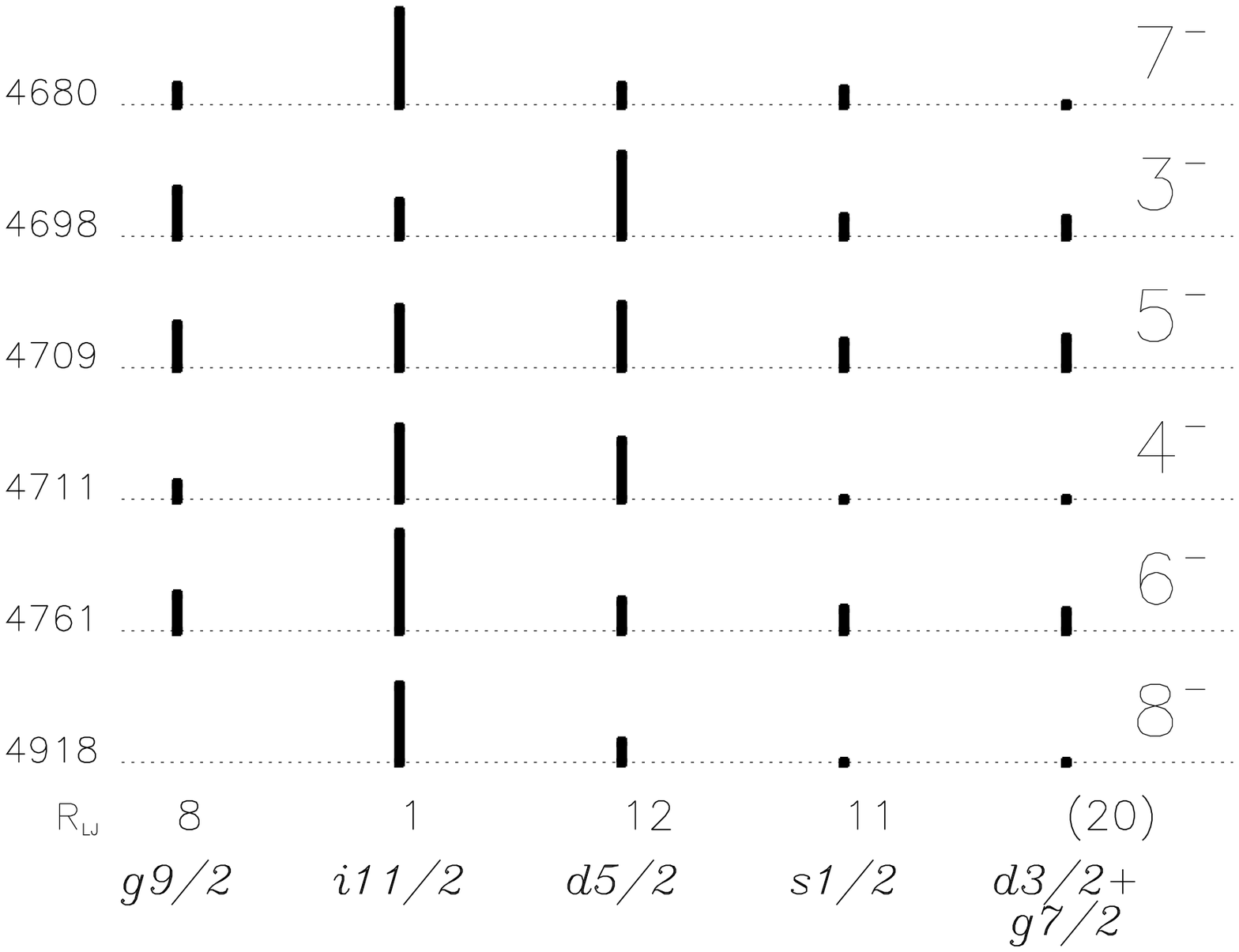}}
}
\resizebox{\hsize}{07.2cm}{
{\includegraphics[angle=00]{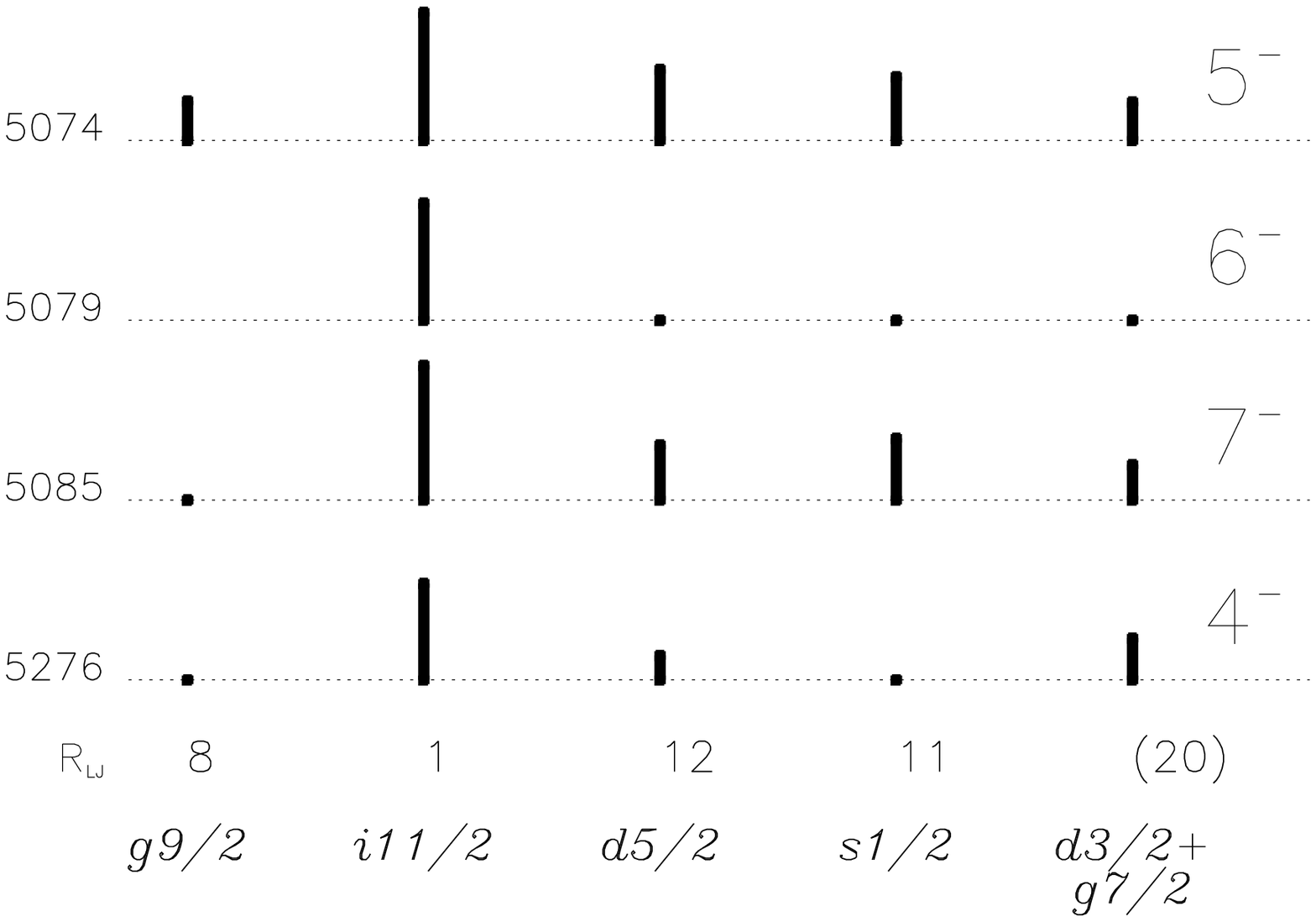}}
}
\end{figure}

In Fig.~\ref{IARschemiE.f5.p3} we show the \excFs\ in a
schematic manner.  For each of the ten states in discussion and for
each IAR the mean cross section $\sigma^{\alpha\,I}_{LJ}$ is shown.
All levels in discussion show a pronounced excitation by the \iEhlb\
IAR. They have weak counterparts on all other IAR.

In reality due to the low \peneTra\ of the \iEhlb\ particle, the cross
sections for the \gNhlb, \dFhlb,\sOhlb\ IAR must be be reduced by the
\peneTra\ ratio $R_{LJ}$=8,~12,~11,~(20) (Eq.~\ref{eq.ratio} and
Tab.~\ref{ratio.spWid}) 
in relation to the \iEhlb\ IAR, see Tab.~\ref{allStates.info}.
Taking into account these values, Fig.~\ref{IARschemiE.f5.p3}
demonstrates that the ten states are rather pure.

\subsection{\AngDs\ of \Pb(p,~p')
}

\begin{figure}[htb]
\caption[
\AngDs\ i11 f5 for 4- 6-
]
{\label{IARangDis.iEf5.45}%
\AngDs\ for the 4.71\,MeV doublet partner with spin $4^{-}$ and  the
state with spin $6^{-}$.
The mean cross section for a pure \cfg\ \iEhlb\fFhlb\ calculated by
Eq.~\ref{eq.avg.c.s} is shown by a dashed line; the corresponding
\angDs\ calculated by Eq.~\ref{eq.diff.c.s} is shown by the dotted
curve. Both calculated curves are corrected for the energy dependent
\peneTra\ by Eq.~\ref{eq.sigmaCorr},~\ref{eq.peneTra}.
}
\resizebox{\hsize}{05.5cm}{
{\includegraphics[angle=00]{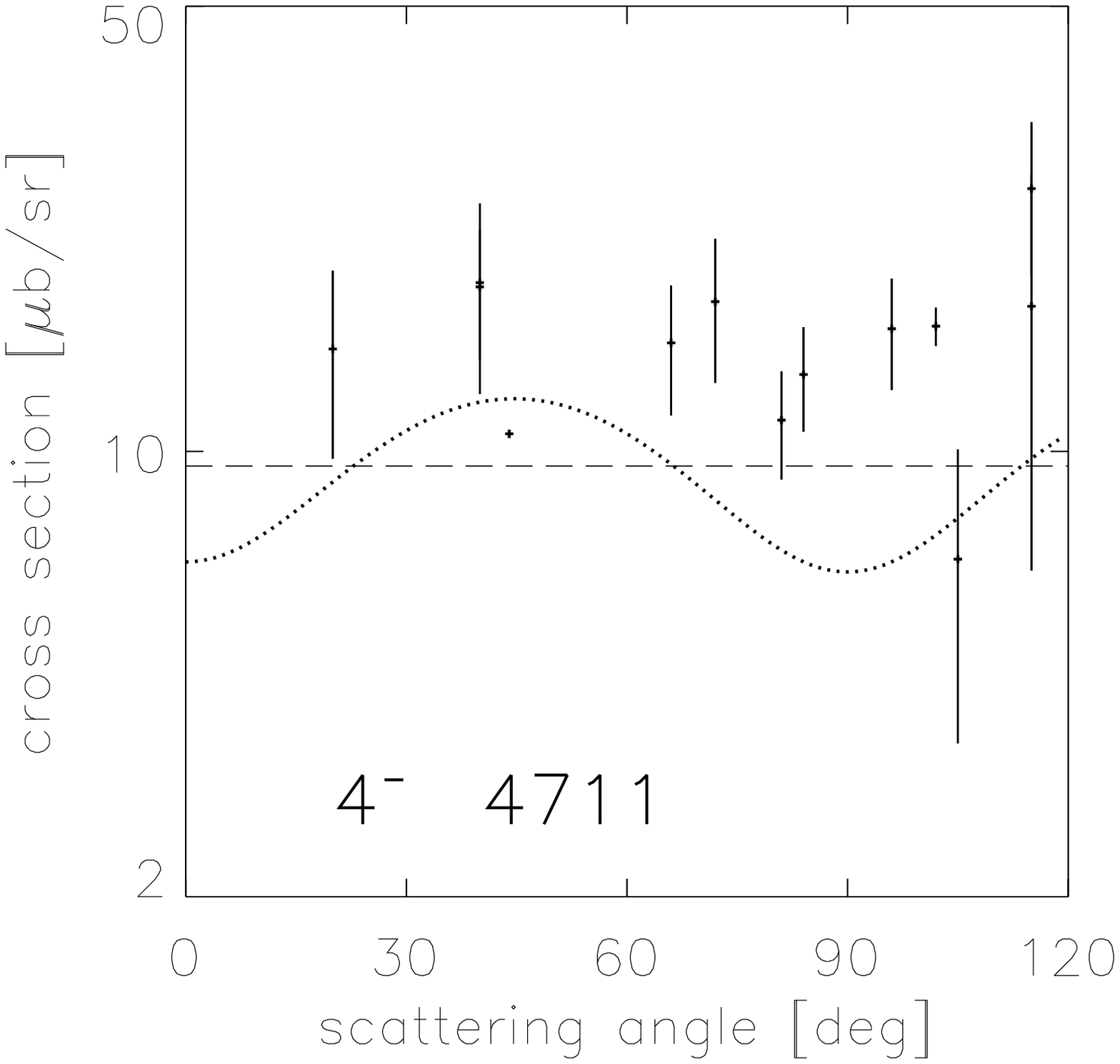}}
{\includegraphics[angle=00]{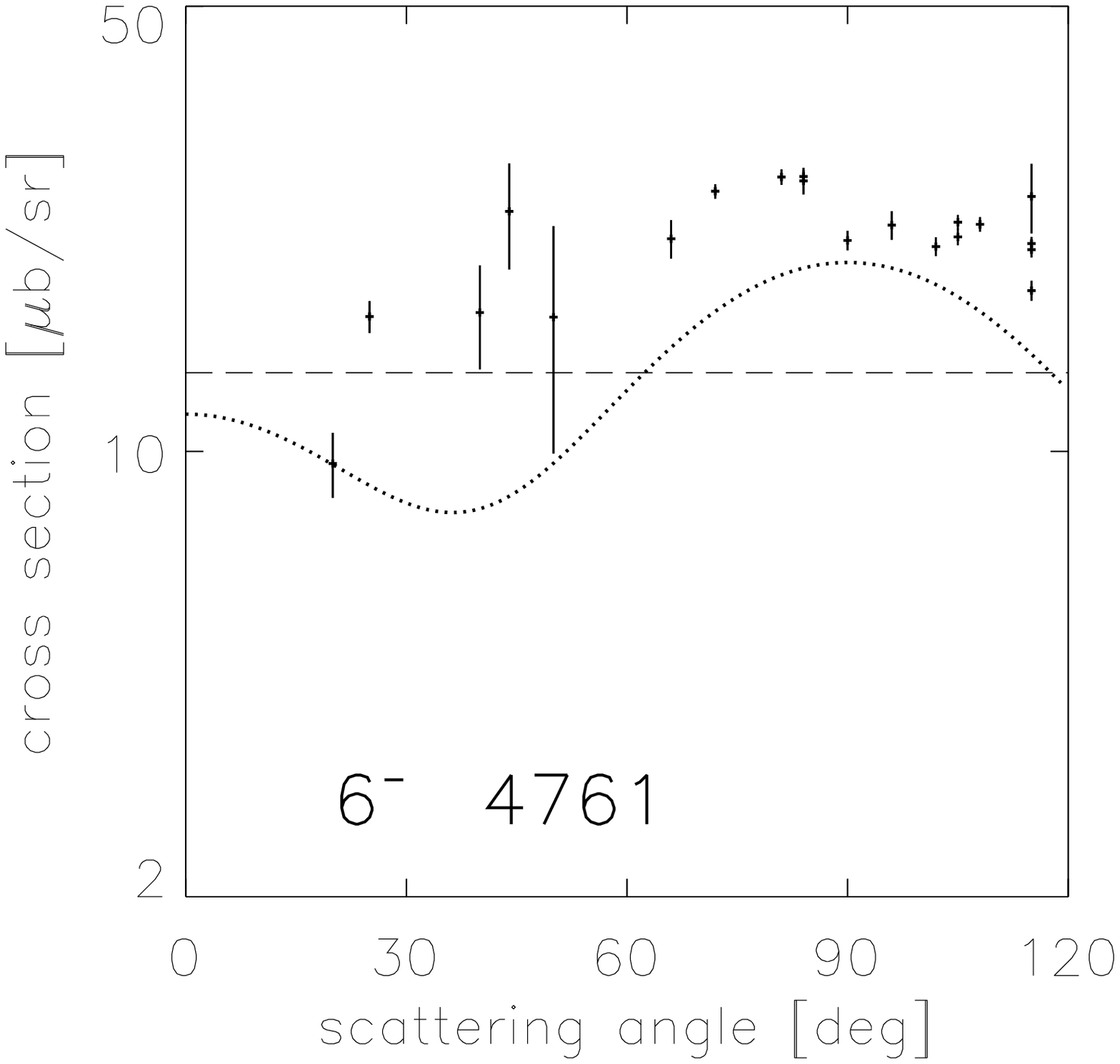}}
}
\end{figure}

\begin{figure}[htb]
\caption[
\AngDs\ i11 f5 for 7- 8-
]
{\label{IARangDis.iEf5.678}%
\AngDs\ for the \iEhlb\fFhlb\ states with spins $7^{-},8^{-}$.
For the 4918 state \cite{1968WH02} 
measured an \excF\ at a scattering angle of $\Theta=158^\circ$. 
The cross section on top of the IAR with $\sigma^{\alpha
I}_{LJ}=20\pm2\mu b/sr$  agrees with the value near $\Theta=22^\circ$
assuming symmetry around $\Theta=90^\circ$. 
For other details see Fig.~\ref{IARangDis.iEf5.45}.
}
\resizebox{\hsize}{05.5cm}{
{\includegraphics[angle=00]{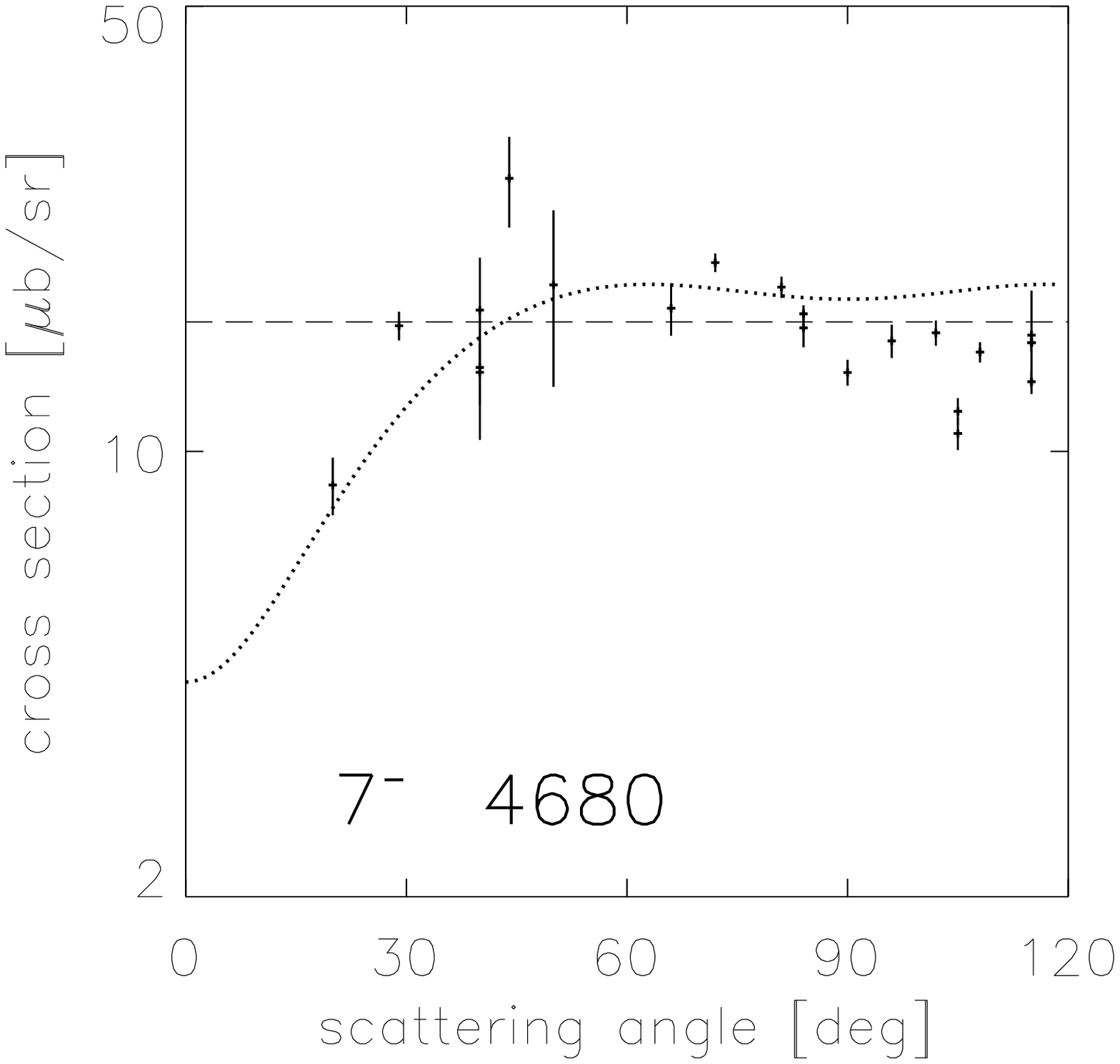}}
{\includegraphics[angle=00]{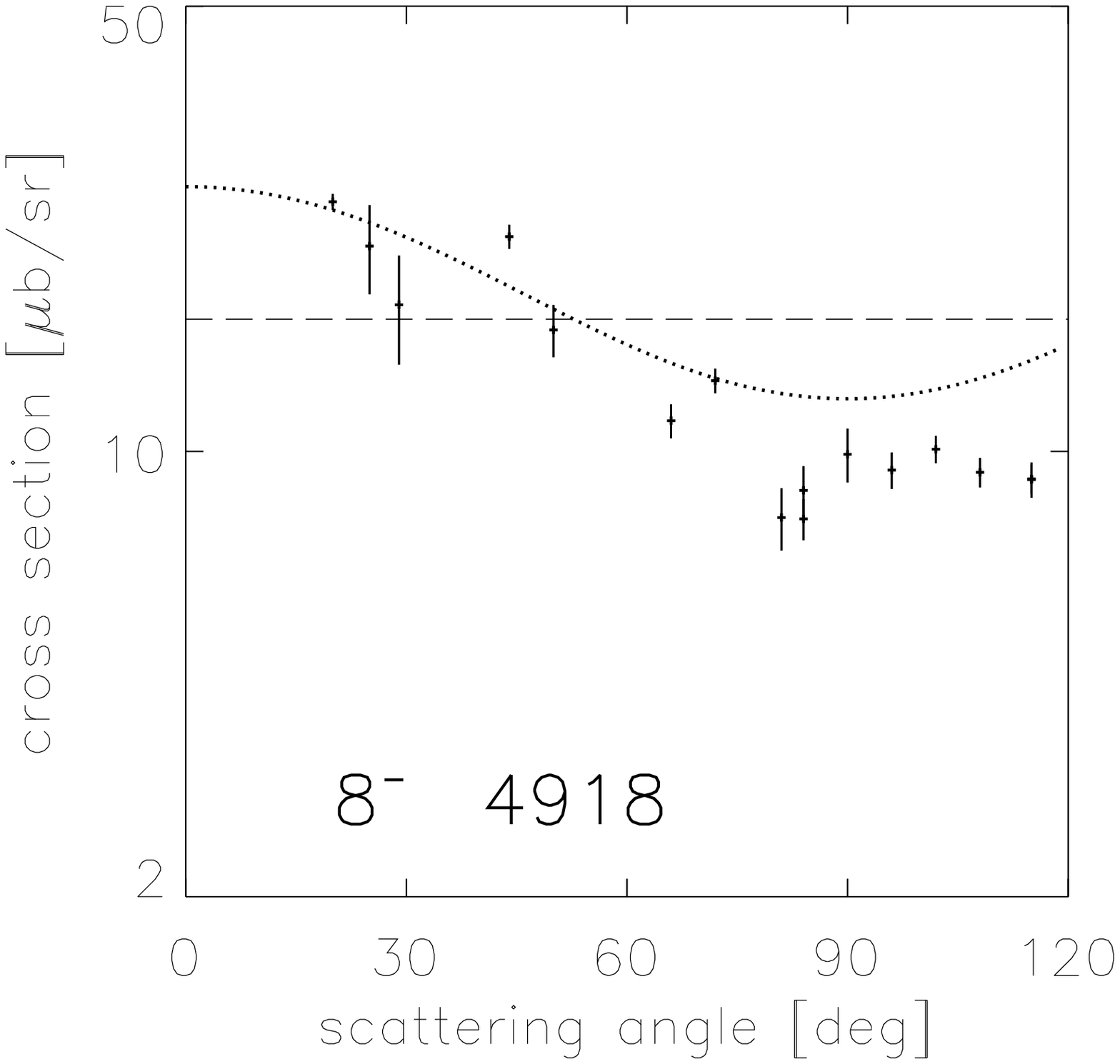}}
}
\end{figure}

\begin{figure}[htb]
\caption[
\AngDs\ i11 p3
]
{\label{IARangDis.iEp3}%
\AngDs\ states for  the \iEhlb\pThlb\ states with spins
$4^{-},5^{-},6^{-},7^{-}$. 
For details see Fig.~\ref{IARangDis.iEf5.45}.
}
\resizebox{\hsize}{05.5cm}{
{\includegraphics[angle=00]{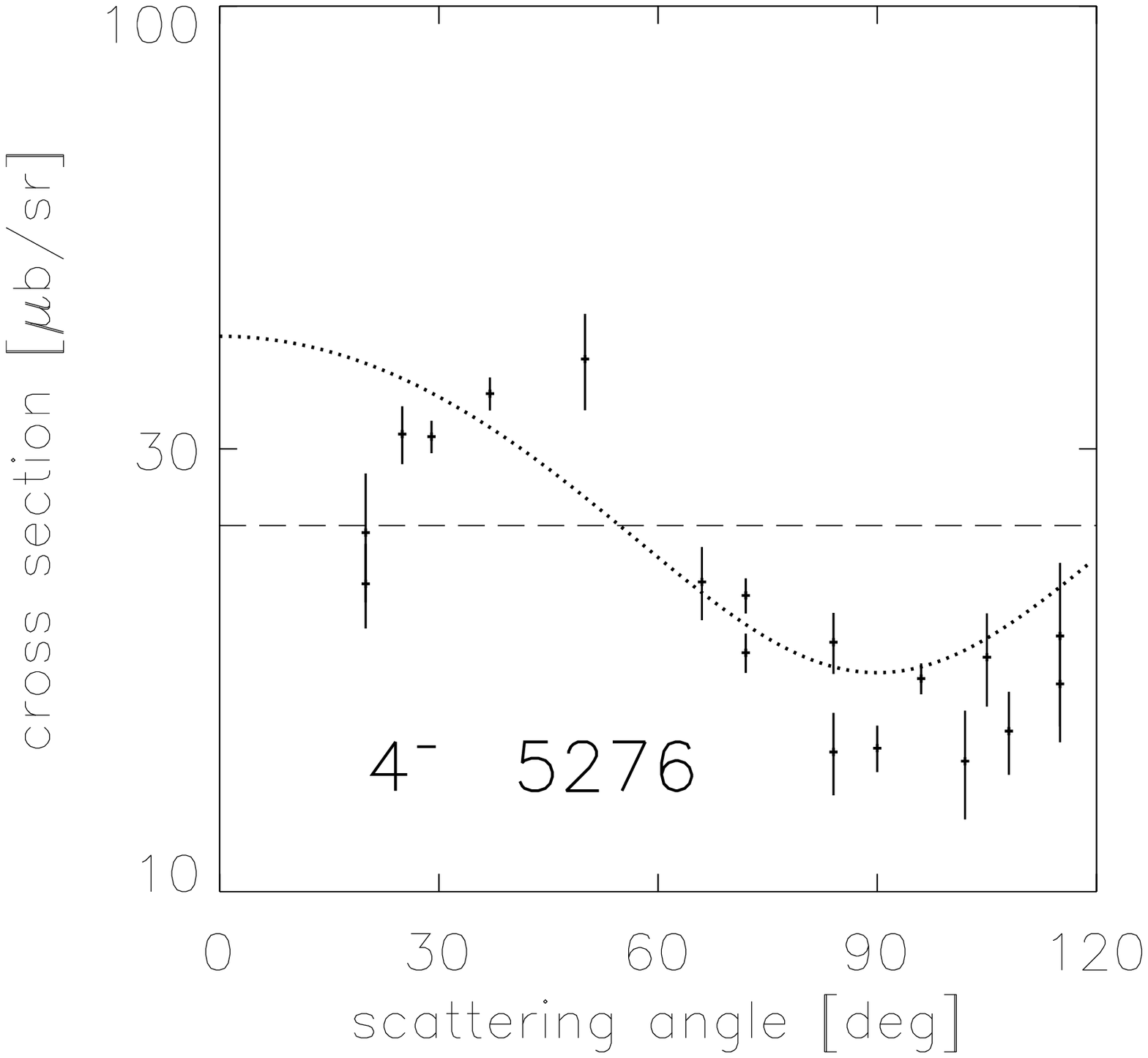}}
{\includegraphics[angle=00]{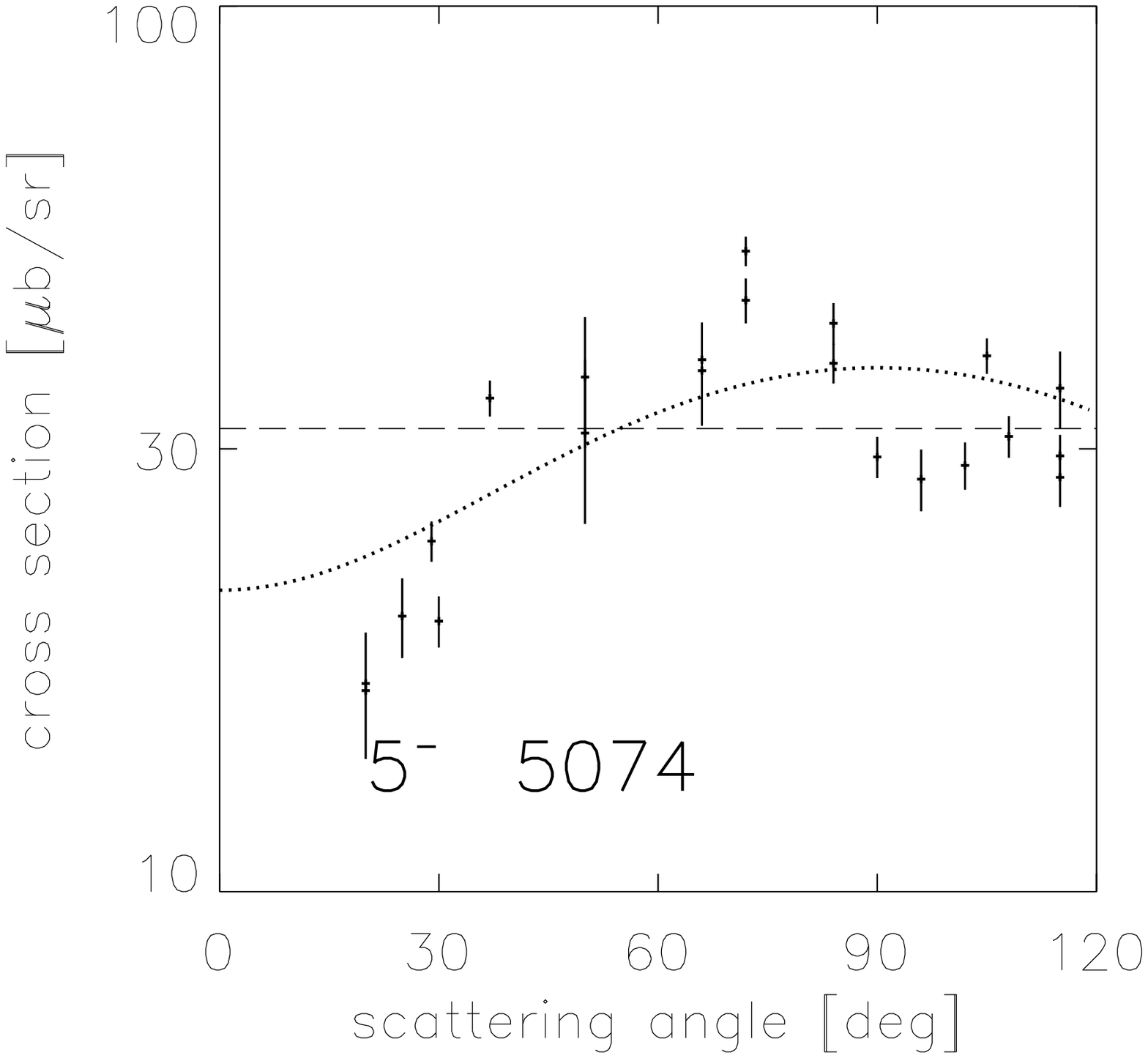}}
}
\resizebox{\hsize}{05.5cm}{
{\includegraphics[angle=00]{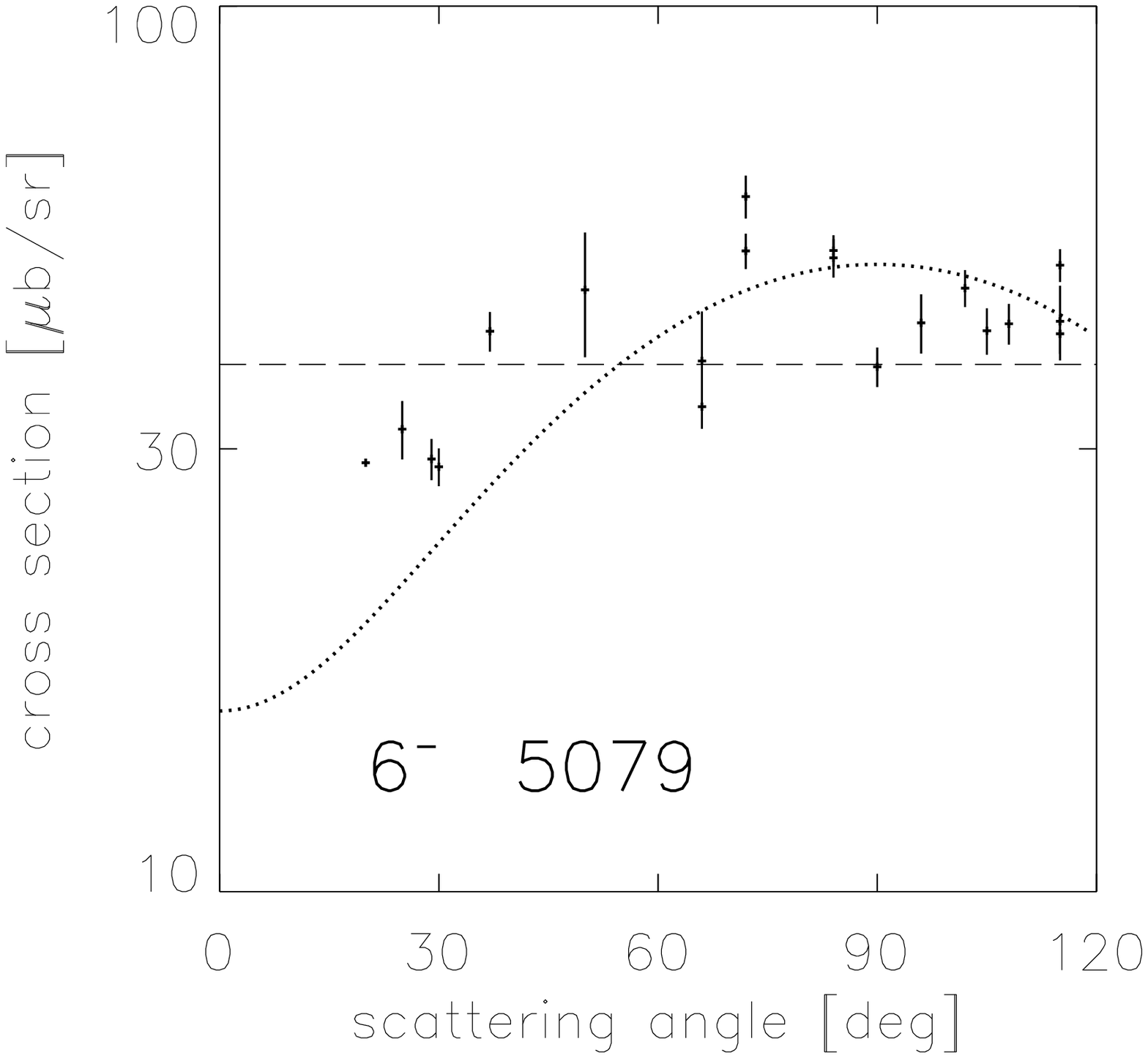}}
{\includegraphics[angle=00]{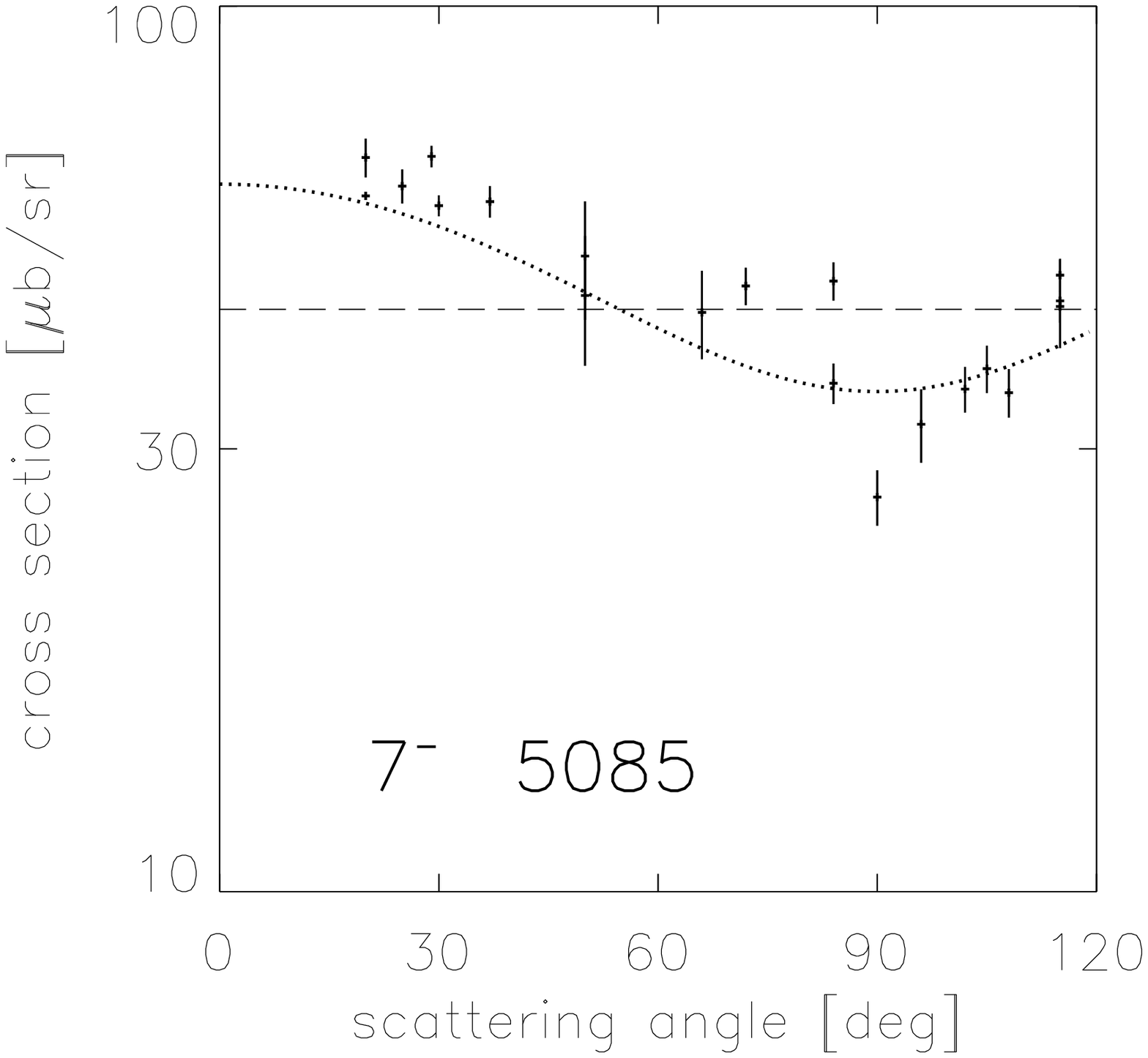}}
}
\end{figure}

In Fig.~%
\ref{IARangDis.iEf5.45}, \ref{IARangDis.iEf5.678}
 and
 \ref{IARangDis.iEp3}
we show the \angDs\ for some members of
the \iEhlb\fFhlb\ multiplet and all members of the \iEhlb\pThlb\
multiplet. 
The cross sections are shown on a logarithmic scale in
$\mu b/sr$; the scale of the
scattering angles is $0^\circ<\Theta<120^\circ$, the highest angle
where we could measure was $115^\circ$; below $20^\circ$ the spectra
became unusable due to increasing slit scattering.
The spin assignment is discussed below. 

Calculations for the pure \ph\ \cfgs\  by Eq.~\ref{eq.diff.c.s} are
inserted for the \angDs\ (dotted line) and the angle averaged (mean)
cross section $\sigma^{\alpha I}_{LJ}$ (dashed line).
The absolute value of the calculated \angDs\ has been adjusted to an
approximate best-fit for the $8^{-}$ state of the \iEhlb\fFhlb\ group
and for the $7^{-}$ state of the \iEhlb\pThlb\ group yielding a more
precise value of $\Gamma^{s.p.}_{\iEhlb}$.
For the states with other spins no adjustment has been done except for
the energy dependence of the \peneTra\ (Eq.~\ref{eq.sigmaCorr},~\ref{eq.peneTra}).

For the \iEhlb\pThlb\ group there is a general agreement of the
mean cross section with the calculation, whereas  for the the
\iEhlb\fFhlb\ group only the states with highest spins $7^{-},8^{-}$
agree  with the expectation of a rather pure \cfg.
The 4698 level has a cross section about ten times higher than expected.
For the  $4^{-},6^{-}$ states
the shape of the \angDs\  agrees with the expectation of a rather pure
\cfg, but the  angle averaged (mean) cross section is around 50\% higher. 
For the $5^{-}$ state  the \angD\ ({\it not shown}) deviates from a
\iEhlb\fFhlb\ distribution at forward angles $\Theta<60^\circ$ up to a
factor~4.

Note that the \angD\ of the members  with the
highest spin $I=J+j$ 
for the \cfgs\ \iEhlb\pThlb\ and \iEhlb\fFhlb\ 
(Fig.~\ref{Pb_pp_50.53}: 5085 $7^{-}$, 
 Fig.~\ref{Pb_pp_46.50}: 4918 $8^{-}$) 
are similar, both show the characteristic minimum at $\Theta=90^\circ$. 
As expected for the lowest spin $I=J-j$, the 5276 $4^{-}$ state
(Fig.~\ref{Pb_pp_50.53}) exhibits the characteristic forward peaking
similar as for the highest spin $I=J+j$.


\subsection{Data from \PbS(d,~p)
}

Tab.~\ref{allStates.info} gives the results from our \PbS(d,~p)
measurement for the ten levels in discussion.
The precision of the excitation energies is slightly better than that 
from the IAR-pp' measurement.
This may be partly explained by satellite lines due to an 
atomic effect which deteriorates the \Pb(p,~p')  but not the
\PbS(d,~p) spectra \cite{LMUrep2003}.
The energy of the 5075 level with a deviation of about
2\,$\sigma$ from \cite{Schr1997} agrees with the result from
the IAR-pp' measurement. 
Some levels have a vanishing \PbS(d,~p) cross section, especially the 4680, 4918,
5085\,keV levels.

In Tab.~\ref{allStates.other.1},~\ref{allStates.other.2} we add the
information derived from the Q3D \expt\ on \PbS(d,~p) for the region
4.5\,MeV$<E_x<$5.3\,MeV for levels {\it not} belonging to the ten
states in discussion.

\section{Results and discussion
}

A key assumption of the shell model is the existence of rather pure
1-particle 1-hole excitations if the spacing of the model \cfgs\ is
higher than the average matrix element of the residual interaction.
We verified this assumption for  multiplets excited by the weakest
positive parity IAR in \Bi. The extremely weak excitation of
the lowest 2-particle 2-hole states (especially the 5239 $0^{+}$
state) adds confidence in the shell model.

In the SSM four multiplets 
\iEhlb\pOhlb,
\iEhlb\fFhlb, 
\iEhlb\pThlb,
\iEhlb\fShlb\ 
are expected to be built with the
$\iEhlb$ particle at energies
$E_x$=4.210, 4.780, 5.108, 6.550\,MeV, respectively, see
Fig.~\ref{IAR.scenario}. 
The goal of this paper is the identification of the \iEhlb\fFhlb,
\iEhlb\pThlb\ neutron \ph\ multiplets;  the states
containing the major strength of the \cfg\ \iEhlb\pOhlb\ are 
known \cite{AB1973,Schr1997,Valn2001}, see  also Tab.~\ref{QfipStri};
for the \iEhlb\fShlb\ group no measurement has been done
(Tab.~\ref{Q3D.params.pp}). 
%

We encounter several problems with the IAR-pp' method,
\begin{itemize}
\item 
the s.p. widths $\Gamma_{lj}^{s.p.}$ for the outgoing particles 
($lj=$\pOhlb, \fFhlb, \pThlb) are  only known to about 10\%,
\item 
the energy dependence of the s.p. widths is rather strong and
its   slopes are not well known. In the region of interest a
systematic error of around 20\% has to be assumed,
\item 
the mean cross section $\sigma^{\alpha I}_{LJ}$ of a state bearing the
main strength of a \cfg\ with 
angular momenta  $l$ is strongly affected by the presence of a slight
admixture of a \cfg\ with lower angular momentum $l-2$ due to the
higher \peneTra,
\item 
the anisotropy of the \angD\ is highly sensitive to the mixture of the
\cfgs. This is especially true for a small admixture of a \cfg\ $|lj>$
with $j=l+1/2$ to a \cfg\ with $j=l-1/2$. In rare cases the
anisotropy \coefs\ $a_K/a_0,\, K=2,4,6,8$ allow to determine the
relative mixing of \cfgs\ $|LJ>\otimes|lj>$ with $l=1,3,5$, $j=l\pm1$.
\item 
the \angD\ of states with natural parity often exhibit strong
forward peaking via the direct-(p,~p') reaction,
\item 
the s.p. widths $\Gamma_{\iEhlb}^{s.p.}$, $\Gamma_{\jFhlb}^{s.p.}$ of
the two weakest IAR are only known to 70\%.
\end{itemize}

\begin{figure}[htb]
\caption[
States excited on the \iEhlb\ IAR
]
{\label{ergbns.centroid}%
In the {\bf upper panel},
the centroid excitation energy (eq.\ref{eq.centroid})
and the total \cfg\ strength $\sum_I
|c^{I}_{LJ,lj}|^2$ are shown.
The centroid energies agree with  energies  $E_x$~= 4.210,
4.780, 5.108\,MeV of the SSM for the three \cfgs\  \iEhlb\pOhlb,
\iEhlb\fFhlb, \iEhlb\pThlb. 
The total \cfg\ strengths are close to unity using the 
s.p. widths  from Tab.~\ref{ratio.spWid}.
In the {\bf  lower panel} the excitation energies $E_x$ and 
the partial strength $|c_{LJ,lj}|^2$
for the  states bearing the main strength of
the \pOhlb, \fFhlb, \pThlb\ \cfgs\   are shown.
The cross sections $\sigma^{\alpha I}_{LJ}$ from
Tab.~\ref{allStates.info} 
are  converted to partial strengths by eq.\ref{eq.avg.c.s}
with s.p. widths from Tab.~\ref{ratio.spWid} and 
corrected for the energy dependence of the \peneTra
(Eq.~\ref{eq.sigmaCorr},~\ref{eq.peneTra}). 
For the \iEhlb\pOhlb\ multiplet the {\it sum} of the partial strengths
of the three $5^{-}$ and the three $6^{-}$ states at
4.0\,MeV$<E_x<$4.5\,MeV is shown.
The value for the 4698\,$3^{-}$ state (left out in determining the
centroid energy) is reduced by a factor~3. Both this $3^{-}$ and the
neighbouring $5^{-}$ state   are affected by the direct-(p,~p')
reaction yielding a much larger value than unity.
The SSM expects a value $|c_{LJ,lj}|^2=1$ (dotted line).
}
\resizebox{\hsize}{12.05cm}{
%
%
{\includegraphics[angle=00]{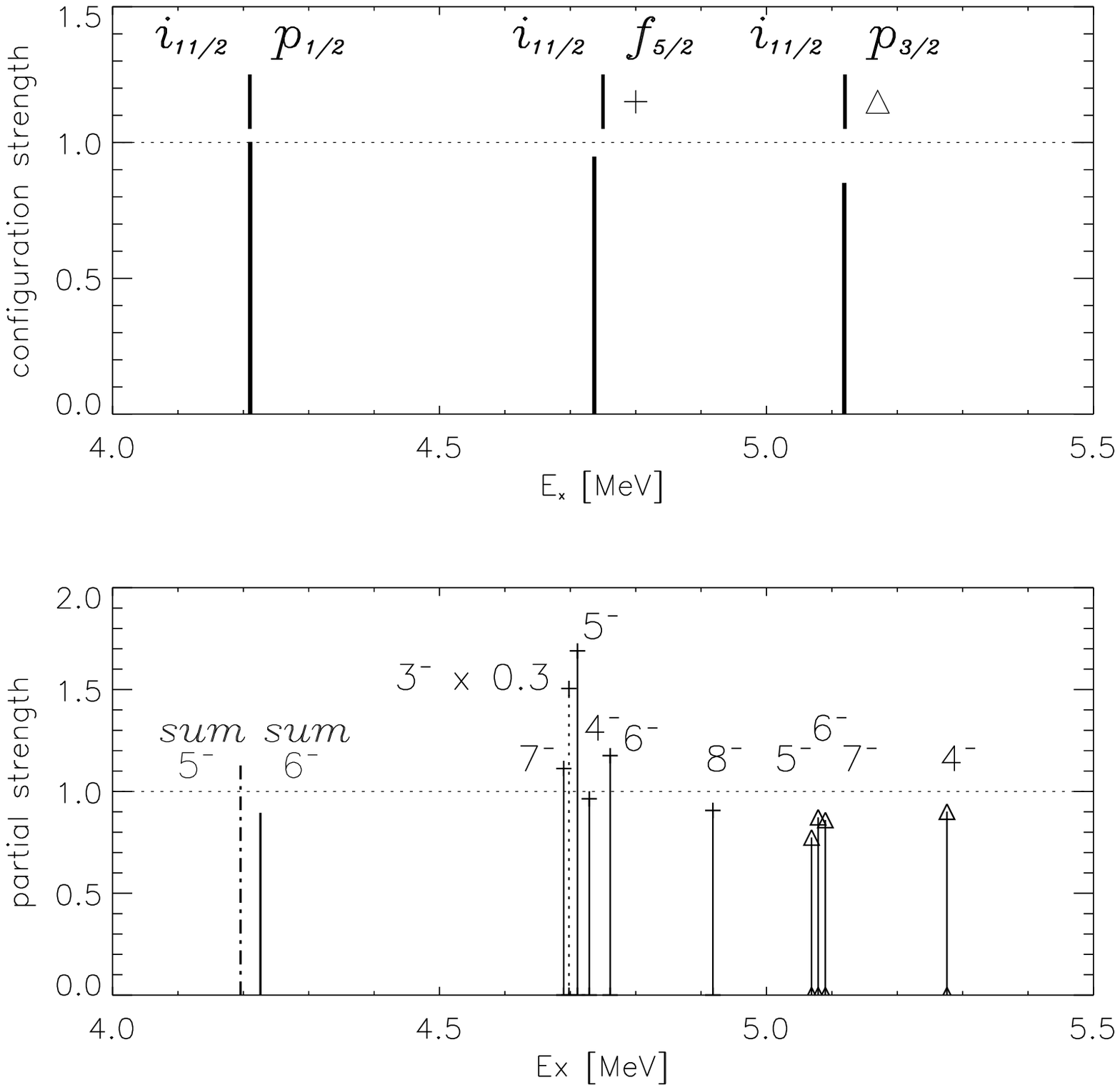}}
}
\end{figure}


\subsection{Centroid Energy}

The  states strongly excited by the \iEhlb\ IAR  can be grouped into
three parts,  the first part  at $E_x\approx\ $\,4.2\,MeV belongs to
the group of states  strongly excited by the \gNhlb\ IAR mainly, 
the second part
at $E_x\approx\ $\,4.6-4.8\,MeV (except for the 4698 $3^{-}$ state)
and
the third part
at $E_x\approx\ $5.1\,MeV are excited by no other IAR  strongly.
The number of states in the second and third group is six and four.
(In the following discussion, the 4698 $3^{-}$ state is  omitted 
since it is affected by a large 
direct (p,~p') contribution starting at least at scattering angles
$E_x<115^\circ$, the maximum angle for the Q3D magnetic spectrograph
in the current shape.)

We derive the centroid energies from the excitation energies $E_x$ 
and the  angle averaged (mean) cross sections $\sigma^{\alpha I}_{LJ}$ 
of the remaining five and four states 
given in Tab.~\ref{allStates.info}.
First, we correct the mean \expt al cross sections by the large change
of the \peneTra\ for the outgoing particles across the range of
excitation energies, 
\begin{eqnarray}
\label{eq.sigmaCorr}
\tilde\sigma^{\alpha I}_{LJ}
=
p^2(E_x^{\alpha I}(LJ))
\sigma^{\alpha I}_{LJ}.
\end{eqnarray}
The energy dependence of the \peneTra\ 
is calculated \cite{1971CL02} and can be linearly approximated by 
\begin{eqnarray}
\label{eq.peneTra}
p(E_x^{\alpha I}(LJ))= 1 + 3.5\,
{\frac{
E_x^{\alpha I}(LJ) - E_{LJ,lj}^{SSM}  }{
E_{LJ,lj}^{SSM}}}.
\end{eqnarray}
The approximation is reasonable near the SSM excitation energy of
the \ph\ \cfgs\  $|LJ>\otimes|lj>$ for all relevant values of $lj$.
(The slope varies between 2.0 and 6.0 for 8\,MeV$<E_{p'}<14$\,MeV and
$l=1,3,5$, for higher $l$-values the slope always becoming steeper.)

We then calculate the centroid energy by the weighted mean 
\begin{eqnarray}
\label{eq.centroid}
<E_x(LJ)> = \sum_{\alpha I} \tilde\sigma^{\alpha I}_{LJ}
E_x^{\alpha I}(LJ) 
\end{eqnarray}
Fig.~\ref{ergbns.centroid} {\it upper panel} shows 
the centroid energies; clearly they coincide  with the prediction of
the SSM model. 
We note that the adjustment of the s.p. widths discussed 
in appendix~C does
not affect the values of the centroid energies much.

The ratio of the sum of the  angle averaged (mean) cross sections
$\tilde\sigma^{\alpha I}_{LJ}$ (converted to \cfg\ 
strengths by use of Eq.~\ref{eq.avg.c.s}) for the groups related to the
\pOhlb, \fFhlb, \pThlb-particle compares well with the calculated
ratio derived from the s.p. widths of Tab.~\ref{ratio.spWid}.
We note that the shown deviations of the \cfg\ strengths from unity
are already lessened by improved s.p. widths as discussed in appendix~C;
using the values from \cite{1968WH02,1969RI10} the deviations are
larger, but still in the range 10-30\%.

Both the agreement of the centroid energies 
and the  approximate agreement of the 
\cfg\ strengths with the SSM expectation 
favour the identification of the states shown in
fig.~\ref{Pb_pp_46.50}-\ref{ergbns.centroid} 
and
Tab.~\ref{allStates.info} as the members of the \iEhlb\fFhlb\ and
\iEhlb\pThlb\ multiplets.

\subsection{
Proton \ph\ \cfgs
}

IAR-pp' is sensitive to neutron \ph\ \cfgs\ only
(Eq.~\ref{eq.IAR.sum}). Yet with robust values of the s.p. widths and
reasonable functions for the energy dependence of the \peneTra, a
missing \cfg\ strength can be determined.  In case the unitarity of
the truncated \cfg\ space can be trusted, by this means even
amplitudes of proton \ph\ \cfgs\ can be determined. An example is give
in appendix~A.

The \cfgs\ \fShlb\sOhlb, \iEhlb\pThlb\ have similar SSM energies
5.011, 5.108 \,MeV, respectively, including the Coulomb shift
$\Delta_C=-0.30\pm0.02$\,MeV (appendix~A).
Hence an admixture of the proton
\ph\ \cfg\ \fShlb\sOhlb\ to the neutron \ph\ multiplet can be
expected for the state with spin $4^{-}$ at $E_x$=5.276\,MeV.
It does not change the \angD\ of IAR-pp', but reduces the mean cross
section only.
We derive an upper limit of 20\% for the \fShlb\sOhlb\ component.

Similarly the \cfg\ \fShlb\dThlb\ with the
SSM energy 5.462\,MeV may change the structure of the states with a
dominant \iEhlb\pThlb\ \cfg.
Evidently the states with spins  $6^{-},7^{-}$ are not affected due to the
high spin; for the states with spins $4^{-},5^{-}$ we derive 
upper limits of 20\% for the \fShlb\dThlb\ component.

The principle of unitarity for a rather complete set of shell model
\cfgs\ can be used to predict one $4^{-}$ states with dominant \cfg\
\fShlb\sOhlb\  in the region $E_x=5.0\pm0.2$\,MeV 
as has been done successfully for the N=82 nucleus \Ce\
\cite{Heu1969,GNPH140CeIII}. 
Several candidates can be found
(Tab.~\ref{allStates.other.1},~\ref{allStates.other.2}). They should be
weak on all IAR eventually except for the \iEhlb\ IAR and have weak
\PbS(d,~p)  and vanishing \Bi\dHe\ cross sections.

\subsection{The \iEhlb\pOhlb\ \ph\ multiplet
}

The \iEhlb\pOhlb\ strength for spin $5^{-}$  is split up into three
fractions, whereas the $6^{-}$ strength is contained in one state
mainly.
The lowest $6^{-}$ state at $E_x$=3919\,keV 
contains less than 1\% of the \iEhlb\pOhlb\ strength as shown
especially by the 
absence of a detectable \PbS(d,~p) cross section both
with the Buechner and the Q3D \expt.
The centroid energies agree well with the prediction by the
SSM model, see Fig.~\ref{ergbns.centroid}.  
The ratio of the total strength for the $5^{-}$ and
$6^{-}$ states  does not relate as expected
from the SSM as 11:13.
This hints to a considerable part of the \iEhlb\pOhlb\
strength in a higher $6^{-}$ state.

Pure \iEhlb\pOhlb\ states should have isotropic \angDs, but all six
\angDs\ deviate from isotropy. 
Small admixtures of other \cfgs\ like \iEhlb\fFhlb, \iEhlb\pThlb,
\iEhlb\fShlb\ may explain the anisotropy. 

For the fit shown in appendix~A only few data for the \cfg\
\iEhlb\pOhlb\ were used, namely the  4206
state to bear the overwhelming strength  and and a roughly equal
partition of the $\iEhlb\pOhlb\, 5^{-}$ strength into the  4125, 4180,
4296 states. 
The mean cross sections now determined more precisely are in general
agreement with the fit. It thus gives confidence into the fitting
procedure. 

The 4206 state has been already identified by \cite{1968WH02} and 
used to determine the total width of the \iEhlb\ IAR, see
Tab.~\ref{ratio.spWid}. 
%

\subsection{The \iEhlb\fFhlb\ \ph\ multiplet
}

Tab.~\ref{allStates.info} (upper part) gives 
the mean cross section for the states
containing the major part of the \iEhlb\fFhlb\ \cfg, see also
Fig.~\ref{ergbns.centroid}.
The states at 4680, 4698, 4709, 4711, 4761\,keV have rather firm spin
assignments with spins $7^{-}$, $3^{-}$, $5^{-}$, $4^{-}$, $6^{-}$
according to \cite{Schr1997}.
These states -- except for the $3^{-}$ state at $E_x$=4698\,keV again
-- have an \angD\ which can be explained by a rather pure
\iEhlb\fFhlb\ \cfg, see Fig.~\ref{IARangDis.iEf5.45},
\ref{IARangDis.iEf5.678}.

We approximated the \angD\ by calculations for the pure \cfg\ with a
common factor.
According to the theory \cite{Heu1969} this 
factor is described by the total width $\Gamma_{\iEhlb}^{tot}$ which has 
been measured by \cite{1968WH02} and the s.p. widths
 $\Gamma_{\iEhlb}^{s.p.}$, 
 $\Gamma_{\fFhlb}^{s.p.}$,
 $\Gamma_{\pThlb}^{s.p.}$ 
determined by \cite{1969RI10}; the energy dependence is calculated
\cite{1971CL02}. 
In total the systematic uncertainty is about 20\%.
%
We used the adjusted values for the s.p. widths
(Tab.~\ref{ratio.spWid}).
We remark that the determination of the
s.p. widths is complicated since they can be determined only as the
product $\Gamma_{LJ}^{s.p.}$ for the IAR and $\Gamma_{lj}^{s.p.}$ for
the outgoing particles, see Eq.~\ref{eq.diff.c.s}; in addition the
energy dependence of the \peneTras\ is not well known.

Since the states with the main \cfg\ \iEhlb\fFhlb\ and spins
$4^{-}$, $5^{-}$, $6^{-}$, $7^{-}$
may mix with the \cfgs\ \iEhlb\pThlb\ which have a much larger
s.p. width  
$\Gamma^{s.p.}$ due to the l=1 wave instead of l=3, even a small
admixture changes the \angD\  much.
Seemingly this is the case for the $4^{-},6^{-}$ states at $E_x$=4711,
4761\,keV, respectively,
see Fig.~\ref{IARangDis.iEf5.45}, \ref{IARangDis.iEf5.678}, 
hence the sum of the mean cross section according to
Eq.~\ref{eq.avg.c.s} is  larger than unity.

{\it (a) 4918 $8^{-}$}. 
The state at $E_x$=4918\,keV is excited by the \iEhlb\ IAR solely.
A detection on the \jFhlb\ and \dFhlb\ IAR yields cross sections a
factor 10 lower, see Fig.~\ref{IARschemiE.f5.p3}.
The \PbS(d,~p) cross  section is vanishing small, see
Tab.~\ref{allStates.info}. 
The agreement of the \angD\  with the calculation for a pure
\iEhlb\fFhlb\ \cfg\ is remarkable,
see
Fig.~\ref{IARangDis.iEf5.678}. 
The slight deviation may be interpreted by an admixture of the
\cfg\ \iEhlb\fShlb.
The absence of a sizable excitation by any other IAR corraborates the
spin assignment.

At a scattering angle of $158^\circ$, the resonance is rising by a
factor 18.0 over the direct background \cite{1968WH02}. This fact
corroborates the assignment of an unnatural parity.

{\it (b) 4680 $7^{-}$}. 
The \angD\ agrees with a pure \iEhlb\fFhlb\ \cfg; the slight deviation
may be explained by an admixture of \iEhlb\fShlb.


{\it (c) 4711 $4^{-}$}. 
A weak \iEhlb\pThlb\ admixture explains the augmented cross section
(Fig.~\ref{IARangDis.iEf5.45}) by the much higher \peneTra\ of the
\pThlb\ particle.  
A weak excitation by the \PbS(d,~p) reaction is consistent with a
sizeable excitation by the \gNhlb\ IAR (Fig.~\ref{IARschemiE.f5.p3}).

{\it (d) 4761 $6^{-}$}. 
A fraction of about 10\% of the $6^{-}$ \iEhlb\pOhlb\ strength in the 
4761\,keV state relieves both the augmented mean cross section
(Fig.~\ref{IARangDis.iEf5.678}) and 
the discrepancy found while discussing the 
\iEhlb\pOhlb\ strength above.
It is consistent with the detected \PbS(d,~p) reaction; both the
Buechner and the Q3D data can be explained by a 0.10$\pm$0.04
\iEhlb\pOhlb\ admixture.


{\it (e) 4709 $5^{-}$}. 
The deviation of the \angD\ at forward angles can be explained by a
direct-(p,~p') component; it is consistent with the assignment of
natural parity. Seemingly an admixture of \gNhlb\pOhlb\ 
or \iEhlb\pOhlb\ is small; it is consistent with the smaller cross
section for \PbS(d,~p) in relation to the 4711 doublet member.


{\it (f) 4698 $3^{-}$}. 
The 4698 $3^{-}$ state  is excited at forward angles
ten times stronger as predicted by the SSM, but 
from the \excFs\ of
ref.\cite{Zai1968} we derive upper limits for the backward angles 
$150^\circ,170^\circ$ which are consistent with the pure \iEhlb\fFhlb\
\cfg\, in contrast to the  forward angles symmetric to $90^\circ$.
The state is also known to have  sizable \gNhlb\pThlb\ and 
\dFhlb\pOhlb\ components \cite{AB1973}.
The excitation by the \PbS(d,~p) reaction is consistent with a rather
strong \dFhlb\pOhlb\ component.

A possible alimentation on top of the \iEhlb\ IAR via the exit
channel \dFhlb\pOhlb\ may change the
\angD\  due to the higher \peneTra\ of the outgoing
\pOhlb\ particle. 
The rather strong direct-(p,~p') component contributes in
addition. Therefore the interpretation of this state is complicate.

The \angD\ for the 4.70\,MeV level shown by
\cite{1968WH02,Zai1968} is 
interpreted incorrectly by the authors. Namely the re\-so\-lu\-tion of 
about 35\,keV was
insufficient to resolve this state from the neighbouring multiplet at
4680, 4709, 4711\,keV.  
So the strong excitation  by the \iEhlb\
IAR is  not due to the excitation of the $3^{-}$ state alone, but at
least equally to the $7^{-}$, $5^{-}$, $4^{-}$ multiplet around it.
So the surprise about the strong excitation of the ``4.692\,MeV''
level \cite{1969RI10}  is solved.

\subsection{The \iEhlb\pThlb\ \ph\ multiplet
}

The \angDs\ of the four states containing most of
the \iEhlb\pThlb\ strength is
shown in Fig.~~\ref{IARangDis.iEp3}. 
Calculations for spins $4^{-}$, $5^{-}$, $6^{-}$, $7^{-}$ are
inserted. 
Tab.~\ref{allStates.info} (lower part) gives 
the mean cross section for the triplet levels at 5075, 5079, 5085
and the 5276 level.
Comparing the cross sections for the resolved triplet levels at
$\Theta=90^\circ,22^\circ$ to the data points at
$\Theta=90^\circ,158^\circ$ (ie. symmetric to $90^\circ$) from the
\excF\ of the 5.071\,MeV level unresolved by \cite{1968WH02} we find
agreement within 10\%. This shows that the direct-(p,~p') contribution
is low.

{\it (a) 5085 $7^{-}$}.
The 5085 state is assumed to have spin $7^{-}$ \cite{Schr1997}.
Its \angD\ is well fitted by assuming  a pure \iEhlb\pThlb\ \cfg\ and
very similar to that from the \iEhlb\fFhlb\ \cfg\ with the highest
spin $I=J+j$.
A preliminary analysis of more data designates the $8^{-}$ member of
the \gNhlb\fShlb\ multiplet as  a state at $E_x$=5936\,keV. It
exhibits a similar steep rise of the \angD\ towards forward angles
indicating a rather pure neutron \ph\ \cfg\ with spin $I=J+j$, too.

{\it (b) 5075 $5^{-}$, 5079 $6^{-}$}.
The 5075, 5079 states of the triplet are assigned spin
$5^{-}$,~$6^{-}$. 
A reverse spin assignment fits worse, since the mean cross section
of the 5079 level is about 20\% higher, see Tab.~\ref{allStates.info}. 
The 5075 and 5085 states are excited sizable on both the \dFhlb\ and
\sOhlb\ IAR (the 5085 state also on the \dThlb+\gShlb\ doublet IAR).
This may be due to a considerable direct-(p,~p') cross sections and
corroborates the assignment of natural parity spins
in contrast to the low cross section of the 5079 state on all other
IAR, see Fig.~\ref{IARschemiE.f5.p3}.

In the \excFs\ for the scattering angles
$90^\circ,158^\circ$ the cross section is rising by a factor 10.0,
13.7 over the direct background, respectively.  
This fact may hint to a small  contribution from direct-(p,~p')
for the $7^{-}$ state.

{\it (c) 5276 $4^{-}$}. 
The missing $4^{-}$ member is identified as the 5276\,keV state. It is
strongly excited by the \iEhlb\ IAR, but only weakly on all other IAR,
see Fig.~\ref{IARschemiE.f5.p3}.
The \angD\ is well described by a  pure \iEhlb\pThlb\ \cfg,
see Fig.~\ref{IARangDis.iEp3}.
The cross section is somewhat higher than expected, see
Fig.~\ref{IARangDis.iEf5.678}. 
Only a complete fit of all levels participating in the \cfg\ mixing
with $|\gNhlb>\otimes|lj>$ and $|\iEhlb>\otimes|lj>$ and another
readjustment of the s.p. widths will solve the problem.

{\it(d)
Information from \PbS(d,~p).
}
The 4680 $7^{-}$, 4918 $8^{-}$, 5085 $7^{-}$ states have
vanishing \PbS(d,~p) cross sections corroborating the spin and \cfg\ assignments.
We explain the excitation by the \PbS(d,~p) reaction 
for the 5075 $5^{-}$  state  by a weak   \iEhlb\pOhlb\ admixture,
for the 5079 $6^{-}$  state  by a weak \iEhlb\pOhlb\ admixture,
for the 5276 $4^{-}$  state  by a weak   \gShlb\pOhlb\ admixture.
The \gShlb\pOhlb\ admixture of the 5276 $4^{-}$  state is corroborated
by the excitation on the \dThlb+\gShlb\ doublet IAR, see Fig.~\ref{IARschemiE.f5.p3}.

\setlength\LTleft{0pt}
\setlength\LTright{0pt}
\begin{longtable*}{@{\extracolsep{05pt}}p{25pt}p{30pt}p{12pt}p{15pt}p{25pt}p{35pt}p{25pt}p{25pt}p{33pt}p{25pt}p{33pt}p{30pt}p{55pt}}
\caption{%
Energies, spins  and cross sections
for the ten states 
in the range 4.6\,MeV$<E_x<$5.3\,MeV
discussed in the text
with \cfgs\ \iEhlb\fFhlb\ (upper part),
\iEhlb\pThlb\ (lower part).
The energies and the mean cross sections $\sigma^{\alpha I}_{LJ}$
are determined from the \Pb(p,~p') \expt.
The cross sections $\tilde\sigma^{\alpha I}_{LJ}$ on top of the
\iEhlb\ IAR are corrected  for the energy dependence of the \peneTra\
by Eq.~\ref{eq.sigmaCorr}, the cross sections on top of
the \gNhlb, \dFhlb\ IAR are reduced by the \peneTra\ ratio
$R_{\gNhlb}, R_{\dFhlb}$=8,~12 (Eq.~\ref{eq.ratio}).
Cross sections  for pure \iEhlb\fFhlb\ and \iEhlb\pThlb\ \cfgs\
are calculated from Eq.~\ref{eq.avg.c.s}
with $|c_{\iEhlb\fFhlb}|^2=1$ respectively $|c_{\iEhlb\pThlb}|^2=1$
using s.p. widths from Tab.~\ref{ratio.spWid}
(col.~11).
From \PbS(d,~p)
performed with the Q3D facility, energies (col.~2) and cross sections
$\sigma(\approx25^\circ)$ are derived, too.
Spins from \cite{Schr1997} and energies from 
\cite{Valn2001,Schr1997,Rad1996}  are given for comparison. 
}
\endfirsthead
\multicolumn{11}{c}{Energies, spins  and cross sections
for the ten states  
continued \dots}\\
\hline
\hline
%
$E_x$  &$E_x$ &spin &spin&$E_x$ &$E_x$ &$E_x$ & $\sigma(25^\circ)$ &
$ {{\sigma^{\alpha I}_{LJ}}/{R_{LJ}}} $&
$ {{\tilde\sigma^{\alpha I}_{LJ}}   } $&
$ {{\sigma^{\alpha I}_{LJ}}         } $&
$ {{\sigma^{\alpha I}_{LJ}}/{R_{LJ}}} $& ~~~remark\\
keV &keV  & & &keV&keV&keV &$\mu b/sr$&$\mu b/sr$&$\mu b/sr$&$\mu b/sr$&$\mu b/sr$ \\
(p,~p') & (d,~p) & & & & & & (d,~p)&(p,~p') &(p,~p')&(p,~p')&(p,~p') \\
&&&& &&& & on \gNhlb&on \iEhlb &on \iEhlb &on \dFhlb \\
\hline
(a) &(a) &(a) & \cite{Schr1997} &\cite{Valn2001}&\cite{Schr1997}&\cite{Rad1996}& (a)&(a) &(a)&calcul.&(a) &(a) this work \\
\hline
\hline
\endhead
\endfoot
\endlastfoot
\multicolumn{11}{c}{}\\
\hline
\hline
$E_x$  &$E_x$ &spin &spin&$E_x$ &$E_x$ &$E_x$ & $\sigma(25^\circ)$ &
$ {{\sigma^{\alpha I}_{LJ}}/{R_{LJ}}} $&
$ {{\tilde\sigma^{\alpha I}_{LJ}}   } $&
$ {{\sigma^{\alpha I}_{LJ}}         } $&
$ {{\sigma^{\alpha I}_{LJ}}/{R_{LJ}}} $& ~~~remark\\
keV &keV  & & &keV&keV&keV &$\mu b/sr$&$\mu b/sr$&$\mu b/sr$&$\mu b/sr$&$\mu b/sr$ \\
(p,~p') & (d,~p) & & & & & & (d,~p)&(p,~p') &(p,~p')&(p,~p')&(p,~p') \\
&&&& &&& & on \gNhlb&on \iEhlb &on \iEhlb &on \dFhlb \\
\hline
(a) &(a) &(a) & \cite{Schr1997} &\cite{Valn2001}&\cite{Schr1997}&\cite{Rad1996}& (a)&(a) &(a)&calcul.&(a) &(a) this work \\
\hline
 4680.3  &  ~~~--  &$7^{-}$ &($7^{-}$) &4680.7   & 4680.310 & --       &$<2$ & 0.2 &       16 &  16.2 & 0.3 \\
 $\pm$0.5 &         &        &          &$\pm$0.5 &$\pm$0.250 &          &     &     &          &       &     \\
4698.5  &4698.40  &$3^{-}$ & $3^{-}$  &4698.4   & 4698.375 & 4697.9   & 800 & 1.2 &       45 (b) &   6.6 & 1.6 & (b)~see~text \\
 $\pm$0.3 &$\pm$0.15 &        &          &$\pm$0.5 &$\pm$0.040 &$\pm$0.1   &     &     &          &       & \\
 4709.4  &  ~~~--  &$5^{-}$ &($5^{-}$) &4709.5   & 4709.409 & --       &  10 & 1.7 &       11 &  10.5 & 1.4 \\
 $\pm$0.8 &         &        &          &$\pm$3.5 &$\pm$0.250 &          &     &     &          &       &     \\
 4711.2  &4711.0   &$4^{-}$ & $4^{-}$  & --      & 4711.300 & --       &  15 & 0.5 &       15 &   8.6 & 0.9 \\
 $\pm$0.8 &$\pm$0.6  &        &          &         &$\pm$0.750 &          &     &     &          &       &     \\
 4761.9  &4762.1   &$6^{-}$ & $6^{-}$  &4761.8   & 4761.800 & --       &   7 & 0.5 &       19 &  12.4 & 0.3 \\
 $\pm$0.5 &$\pm$0.4  &        &          &$\pm$0.5 &$\pm$0.250 &          &     &     &          &       &     \\
 4918.8  &  ~~~--  &$8^{-}$ &          &4917.6   & --       & --       &$<2$ & 0.1 &       15 &  16.2 (c) & 0.2 & (c) adapted,\\
 $\pm$0.4 &         &        &          &$\pm$1.5 &          &          &     &     &          &       &         & see text\\
\hline									                     
 5074.6   &5074.8  &$5^{-}$ &  --      &5073.7   & 5075.800 & --       &   9 & 0.6 &  32      &  34   & 0.9  \\ 
 $\pm$ .5 &$\pm$0.4 &        &          &$\pm$1.5 &$\pm$0.200 &          &     &     &          &       &     \\
 5079.8   &5079.8  &$6^{-}$ &          & --      & --       & --       &   5 & 0.5 &  40      &  40   & 0.5 \\
 $\pm$ .6 &$\pm$0.7 &        &          &         &          &          &     &     &          &       &     \\
 5085.3   & ~~~--  &$7^{-}$ &($7^{-}$) &5084.7   & 5085.550 & 5085.7   &$<2$ & 0.5 &  46      &  46 (d)   &  1.0 &(d) adapted,  \\
 $\pm$ .4 &        &        &          &$\pm$1.5 &$\pm$0.250 &$\pm$0.2   &     &     &          &       &          & see text\\
 5276.4   &5276.2  &$4^{-}$ &          &5277.1   & --       & --       &  70 & 0.3 &  26      &  27   & 0.5 \\
 $\pm$ .5 &$\pm$0.2 &        &          &$\pm$1.5 &          &          &     &     &          &       &     \\
\hline
\hline
\label{allStates.info}%
\end{longtable*}


\section{Conclusion
}

The shell model is verified to explain the structure of  30
negative parity states  in \Pb\ below $E_x$=5.3\,MeV by 1-particle
1-hole \cfgs, 10 states more than in the first derivation of matrix
elements of the residual interaction by \cite{AB1973}.
The states containing the major strength of the \cfg\
\iEhlb\pOhlb\ were  measured. 
Amplitudes of the \cfg\ \iEhlb\pOhlb\ obtained by 
an update of the fit done by \cite{AB1973} are verified to be
approximately correct, see appendix~A.

The ten states containing the major strength of
the multiplets \iEhlb\fFhlb\ and \iEhlb\pThlb\ are identified.
All states have one dominant shell model neutron \ph\ \cfg\ 
except  for the \iEhlb\fFhlb\ member with the lowest spin $3^{-}$. 
Some minor admixtures of other \cfgs\  derived from the analysis of
\Pb(p,~p') are consistent with results from \PbS(d,~p).
The detection of the members with the highest spins from the
\iEhlb\fFhlb\ and \iEhlb\pThlb\ group (and of \gNhlb\fShlb, too) gives
hope to find even the \iEhlb\fShlb\ group members with spins $8^{-}$
and $9^{-}$ expected at an energy $E_x=$6.550\,MeV which might be
rather pure.

The clear identification of the \iEhlb\fFhlb\ and \iEhlb\pThlb\ group
and the more carefully measured \angDs\ of the states containing
\iEhlb\pOhlb\ strength will help to refine the derivation of a
more complete shell model transition matrix extending the \cfg\ space
up to $E_x\approx\ $5.2\,MeV similar as done by \cite{AB1973}, at
least for the higher spins. 
Admixtures of proton \ph\ \cfgs\ 
can be  obtained in principle by the assumption of a rather complete
subshell closure and the observation of the \ONrule\ and sum-rule
relations; this fact becomes more relevant  since for the higher
proton \ph\ \cfgs\ there is no target for transfer reactions
of the type \Bi\dHe.

The goal to determine of matrix elements of the effective residual
interaction among \ph\ \cfgs\ in \Pb\ can be approached better than
done by \cite{AB1973} due to the higher quality and larger amount of
\expt al data.
The evaluation of existing data from our
measurements of \Pb(p,~p') and \PbS(d,~p) will allow to extend the
\cfg\ space up to eventually $E_x\approx$\,6.1\,MeV. 

\setlength\LTleft{0000pt}
\setlength\LTright{000pt}
\begin{longtable*}{@{\extracolsep{05pt}}p{35pt}p{25pt}p{35pt}p{35pt}p{35pt}p{35pt}p{35pt}p{35pt}p{35pt}p{35pt}p{35pt}}
\caption{Update of the fit from \cite{AB1973}:
Unitary transformation 
$||c_{LJ,lj}^{\alpha\,I,(\nu,\pi)}||$
of the  shell model \ph\ \cfgs\ below $E_x=$\,4.5\,MeV 
to the states $|\alpha\,I>$ with spins $4^{-},5^{-},6^{-}$.
The errors of the amplitudes are of the order of 0.01 for amplitudes
close to $|c|=1$  and up to 0.20 for $|c|\approx0$.
\quad {\it Footnote: (a)} A sizable admixture in the order of about
$|c|=0.1$ of \gNhlb\hNhlb\  
and \gNhlb\hEhlb\ is needed to 
fit the \angD; it depends on the component $a_8$ from the fit 
of the \angD\ by 
$ d\sigma(\Theta) /d\Omega = \sum_K^{0,2,4,6,8} a_K P_K(cos(\Theta))$.
}
\endfirsthead
\multicolumn{09}{d}{Update\ of\ fit\  continued}\\
\hline
\hline
$E_x$ & spin& \gNhlb\pOhlb &\gNhlb\fFhlb &\gNhlb\pThlb &\gNhlb\fShlb & 
\iEhlb\pOhlb &
\hNhlb\sOhlb &\hNhlb\dThlb \\
\hline
\endhead
\endfoot
\endlastfoot
\multicolumn{09}{d}{}\\
\hline
\hline
$E_x(keV)$ & spin& \gNhlb\pOhlb &\gNhlb\fFhlb &\gNhlb\pThlb &\gNhlb\fShlb &
\iEhlb\pOhlb &
\hNhlb\sOhlb &\hNhlb\dThlb \\
\hline
3475 &$4^{-}$ &   +.985&  +.060&  --.280&   --.013& &   +.050&  --.176\\
3946 &$4^{-}$ &  --.050&  +.293&  --.030&   ~~.000& &   +.937&   +.118\\
3995 &$4^{-}$ &  --.110&  +.984&  --.065&   ~~.000& &  --.389&   +.018\\
4262 &$4^{-}$ &   +.050&  +.070&   +.569&   ~~.000& &   +.182&   +.559\\
4383 &$4^{-}$ &   ~~.000&  +.037&   +.863&  ~~.000& &   +.100&  --.349\\
\hline
3192 &$5^{-}$ &  +.780&   +.350&   +.220&   --.100&  --.150&  --.230&   +.170\\
3709 &$5^{-}$ & --.430&   +.491&  --.130&   ~~.000&  --.300&  --.520&   +.400\\
3960 &$5^{-}$ & --.010&   +.690&   ~~.000&  ~~.000&  ~~.000&   +.720&   ~~.000\\
4125 &$5^{-}$ & --.050&  --.340&   +.270&   ~~.000&  --.440&   +.330&   +.610\\
4180 &$5^{-}$ & ~~.000&   +.015&   +.590&   ~~.000&   +.720&  --.130&   +.250\\
4292 &$5^{-}$ & --.050&   +.150&   +.620&   ~~.000&  --.400&  --.100&  --.600\\
\hline
3919 ${}^a$ &$6^{-}$ &&  +.981  &  --.062 &   +.119& +.110 &&   +.137 \\
4206        &$6^{-}$ && --.222  &   +.045 &  --.010& +.960 &&  --.326 \\
4383        &$6^{-}$ && --.207  &  --.338 &  ~~.000& +.235 &&   +.900 \\
4480        &$6^{-}$ && --.083  &   +.905 &  --.032& +.231 &&   +.401 \\
\hline
\hline
\label{QfipStri}%
\end{longtable*}


{\appendix{
\section{Update of the fit of states in \Pb\ below $E_x$=4.5\,MeV from IAR-pp' data
}

Tab.~\ref{QfipStri} presents the update of the fit shown in
\cite{AB1973} using the same data base, namely \angD\ measurements of
\Pb(p,~p') near the \gNhlb\ IAR \cite{HJG1972} (see also an edited
version in 
\cite{AHwwwHOME}), \PbS(d,~p) data from the Buechner \expt\ as
described in the text,  and other data from \cite{NDS1971}, but in
addition   data from the \Bi\dHe\ \expt\ \cite{Mai1983} done in 1981. 

The results for spins $2^{-},3^{-},7^{-}$ are only slightly changed by
a readjustment of the s.p. widths and therefore not shown here.
The errors are not shown, but of the order of 0.01 for
amplitudes close to $|c|=1$  up to 0.20 for $|c|\approx0$.
The detailled discussion is complicated, but foreseen to be done in
a future publication.
The errors are similar to those in 
\cite{AB1973} since the data base is identic except for the new
\Bi\dHe\ data. The essential difference of the update from
\cite{AB1973} consists only in a few other state identifications and
some interchanged spin assignments.

The centroid energies derived from this fit agree well with the
predictions of the SSM  (Tab.~\ref{g9.centroid}). For the proton \ph\
\cfgs\ we derive a Coulomb shift of $\Delta_C=-0.30\pm.02$\,MeV.
}

\section{$W$-, $\bar Z$- and anisotropy \coefs\
}

For the programming of Eq.~\ref{eq.diff.c.s} in modern computer
languages we give the definition of the $W$- and $\bar Z$-\coefs\ in
terms of 3j- and 6j-symbols.
The $W$-\coef\ is derived from the 6j-symbol as \cite{1952BB,Edm1957} 
\begin{eqnarray}
\label{eq.W}
W(j_1 j_2 l_2 l_1; j_3 l_3)= (-1)^{j_1 + j_2 + l_2 + l_1}
\left\{ 
\!\!
\begin{array}{c}
{j_1 j_2 j_3}\\
{l_1 l_2 l_3}\\
\end{array}
\!\!
\right\}
.
\end{eqnarray}

The $\bar Z$-\coef\ is converted from the definition
by \cite{1952BB,Edm1957} with $W$- and Clebsch-Gordon \coefs\ as
\begin{eqnarray}
\label{eq.barZ}
\bar Z(abcd;ef)=
\sqrt{(2a+1)(2b+1)(2c+1)(2d+1)}
\nonumber\\
(-1)^{a-c}\sqrt{2f+1}
\left(
\!\!
\begin{array}{ccc}
{a c f}\\
{0 0 0}\\
\end{array}
\!\!
\right)
\left\{ 
\!\!
\begin{array}{ccc}
{a b e}\\
{d c f}\\
\end{array}
\!\!
\right\}
i^{f-a+c}.
\end{eqnarray}


Examples of the anisotropy \coefs\ $a_K/a_0$ for pure \ph\ \cfgs\
$|LJ>\otimes|lj>$ calculated from Eq.\ref{eq.avg.c.s} are shown in
Fig.~\ref{aniso}.
They vary in a systematic manner with the angular momentum and spin
of the particle $LJ$  and the hole $lj$. 
For the lowest and highest spin $I=J\pm j$ the \angD\ has a deep
minimum at $\Theta=90^\circ$ as can be derived from the values of the
anisotropy \coefs\ $a_K/a_0$. 

\begin{figure}[htb]
\caption[
IAR.scenario
]
{\label{aniso}%
Anisotropy \coefs\ $a_2/a_0,a_4/a_0,$ 
of the \angD\ for the \ph\ \cfgs\ 
$|LJ>\otimes|lj>$ with a
\gNhlb, \iEhlb, \jFhlb\ particle coupled to a \fFhlb\ hole. 
The values are connected from the 
lowest to the highest spin $I=J\pm j$.
The \cfg\ \iEhlb\pThlb\  with spins
$I^\pi=4^{-},5^{-},6^{-},7^{-}$ 
(marked explicitly) has $a_4=0$.
}
\resizebox{\hsize}{06.50cm}{
{\includegraphics[angle=00]{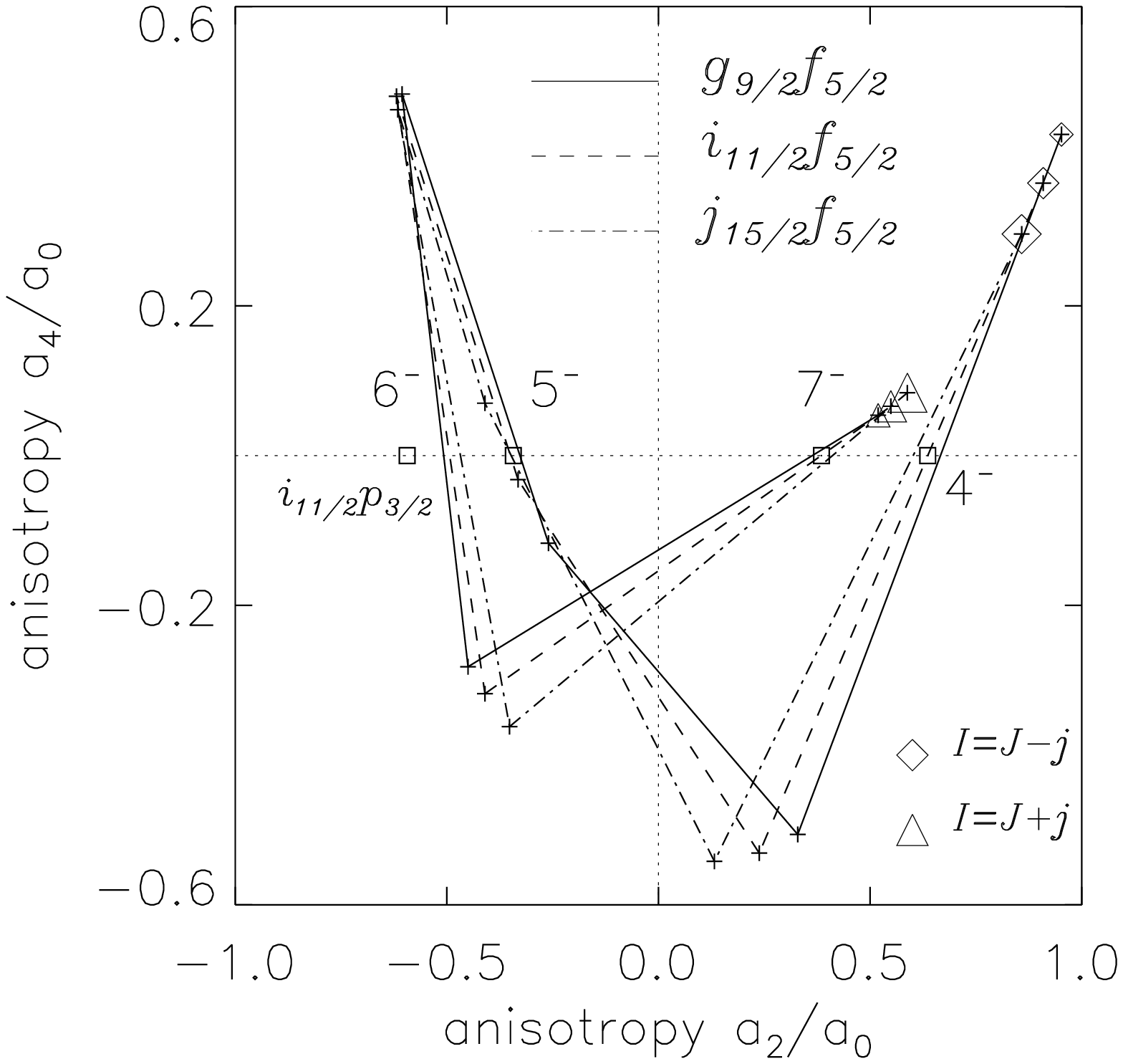}}
}
\end{figure}

\begin{table}[htb]
\caption[g9.centroid
]
{\label{g9.centroid}%
Centroid energies derived from the fit (Tab.~\ref{QfipStri}).
}
\begin{tabular}{|c|cc c| c|cc |}
\hline
&\gNhlb\pOhlb &\gNhlb\fFhlb &\gNhlb\pThlb &
\iEhlb\pOhlb &
\hNhlb\sOhlb &\hNhlb\dThlb \\
 & MeV& MeV& MeV& MeV& MeV& MeV\\
\hline
SSM & 3.41 & 3.98 & 4.31 & 4.21 & 4.21 & 4.56 \\
fit  & 3.40 & 3.93 & 4.32 & 4.18 & 3.92 & 4.25 \\
\hline
\end{tabular}
\end{table}

\section{Improved s.p. widths
}

The values of the s.p. widths $\Gamma^{s.p.}_{LJ}$  for the IAR $LJ$
in Tab.~\ref{ratio.spWid} were derived by \cite{1968WH02} from
the analysis of the \excFs.
The values of the s.p. widths $\Gamma^{s.p.}_{lj}$  
for the outgoing particles $lj$ were derived 
by \cite{1969RI10}. They divided the levels 
below $E_x\approx\ $4.5\,MeV excited strongly by the \gNhlb\ IAR into
three groups 
assigned to bear the main strength of the \cfgs\ \gNhlb\pOhlb,
\gNhlb\fFhlb, \gNhlb\pThlb. 
This is good to first approximation (see Tab.~\ref{QfipStri} in
the appendix~A). 
As we have detected similar groups for the \iEhlb\ IAR
we can improve the values. 

Using the s.p. widths $\Gamma^{s.p.}_{lj}$ from
Tab.~\ref{ratio.spWid},
the comparison of the mean cross sections for the states
$E_x<$ 4.5\,MeV mainly excited by the \gNhlb\ IAR 
to the states
4.5\,MeV$<E_x<$5.3\,MeV almost solely excited by the \iEhlb\ IAR  allows to
derive an improved value of $\Gamma^{s.p.}_{\iEhlb}$. 
We find a value of 2.2\,keV with an error of about 10\%, an improvement
from \cite{1968WH02,1969RI10} by a factor~5.

More precise values of $\Gamma^{s.p.}_{lj}$ can be derived by
considering states which  contain essentially only one SSM \cfg. 
With the \gNhlb\ particle there are four states, 
with the \iEhlb\ particle there are six states; in addition the sum of
the \cfg\ strength for the \iEhlb\pOhlb\ \cfg\ is known due to the
analysis of \cite{AB1973} (see Tab.~\ref{QfipStri}) together with our
new Q3D data.

The lowest $4^{-}$ state at $E_x$=3.475\,MeV is a rather pure
\gNhlb\pOhlb\ \cfg\ with 
a considerable \gNhlb\pThlb\ admixture \cite{1968BO16} and
less than 1\% admixture from \gNhlb\fFhlb, \gNhlb\fShlb, see
Tab.~\ref{QfipStri}. 
The 4480 $6^{-}$ state bears the main part of \gNhlb\pThlb\ \cfg\ with
about 20\% admixture from other \cfgs\ not related to the
\gNhlb\ IAR. 
The 3919 $6^{-}$ state bears the main part of \gNhlb\fFhlb\ \cfg\ with
about  1\% admixture of \gNhlb\pThlb; similarly
the 3995 $4^{-}$ state bears the main part of \gNhlb\fFhlb\ \cfg\ with
1.5\% admixture from \gNhlb\pOhlb, \gNhlb\pThlb. 
Due to the much higher \peneTra\ of the $l=1$ particle in relation to
the $l=3$ particle,  in effect the 1-2\% admixture changes the mean
cross section by about 15\%.

Fig.~\ref{ergbns.centroid} shows the partial strength for the ten
states with a \iEhlb\fFhlb\ or \iEhlb\pThlb\ component
in the upper panel. The sum of the values for the 
states with spins  $5^{-},6^{-}$ bearing the essential \iEhlb\pOhlb\
strength are shown, too. 

We adjusted the s.p. widths $\Gamma^{s.p.}_{lj}$ thus that a unitarity
close to unity is obtained simultaneously for the above mentioned
states strongly excited by the \gNhlb\ IAR (3475, 3919, 3995, 4480), 
and the states strongly excited by the \iEhlb\ IAR.
Here we excepted the states 4698, 4709, 4711, 4760 (see discussion
below) and included the sum of the \iEhlb\pOhlb\ \cfg\ strength
obtained.  

The systematic error in the absolute cross sections is around 10\%, so
we stick to the value $\Gamma^{s.p.}_{\pOhlb}$ given by
\cite{1968WH02}, but derive a slightly 
lower value $\Gamma^{s.p.}_{\pThlb}$=14.6\,keV at an
energy $E_{p'}$=10.59\,MeV, and a considerable higher value 
$\Gamma^{s.p.}_{\fFhlb}$=5.2\,keV at an
energy $E_{p'}$=10.92\,MeV, see Tab.~\ref{ratio.spWid}.

{\it Other s.p. widths}.  From a preliminary analysis of some states,
in relation to other IAR rather strongly excited on the \jFhlb,
\gNhlb\ IAR, we derive values for $\Gamma^{s.p.}_{\jFhlb}$,
$\Gamma^{s.p.}_{\fShlb}$ in a similar manner (Tab.~\ref{ratio.spWid}).



\begin{thebibliography}{10}

\bibitem{Yates96c}
{Minfang Yeh, P. E. Garrett, C. A. McGrath, S. W. Yates, and T. Belgya}.
\newblock {\em {Phys.\ Rev.\ Lett.}}, {76}:{1208}, {(1996)}.

\bibitem{Valn2001}
{B. D. Valnion, V. Yu. Ponomarev, Y. Eisermann, A. Goll\-witzer, R.
  Hertenberger, A. Metz, P. Schiemenz, and G. Graw}.
\newblock {\em {Phys.\ Rev.}}, {C63}:{024318}, {(2001)}.

\bibitem{Rad1996}
{E. Radermacher, M. Wilhelm, S. Albers, J. Eberth, N. Ni\-co\-lay, H.-G.
  Thomas, H. Tiesler, P. von Brentano, R. Schwengner, S. Skoda, G. Winter, and
  K. H. Maier}.
\newblock {\em {Nucl.\ Phys.}}, {A597}:{408}, {(1996)}.

\bibitem{Schr1997}
{M. Schramm, K. H. Maier, M. Rejmund, L. D. Wood, N. Roy, A. Kuhnert, A.
  Aprahamian, J. Becker, M. Brinkman, D. J. Decman, E. A. Henry, R. Hoff, D.
  Manatt, L. G. Mann, R. A. Meyer, W. Stoeffl, G. L. Struble, T.-F. Wang}.
\newblock {\em {Phys.\ Rev.}}, {C56}:{1320}, {(1997)}.

\bibitem{NDS1971}
{M. B. Lewis}.
\newblock {\em {Nucl.\ Data Sheets}}, {5}:{243}, {(1971)}.

\bibitem{NDS1986}
{M. J. Martin}.
\newblock {\em {Nucl.\ Data Sheets}}, {47}:{797}, {(1986)}.

\bibitem{1967MO25}
{C. F. Moore, J. G. Kulleck, P. von Brentano, F. Rickey}.
\newblock {\em Phys.\ Rev.}, 164:1559, {(1967)}.

\bibitem{1967RI13}
{P. Rich\-ard, W. G. Weitkamp, W. Whar\-ton, H. Wieman, P. von Brentano}.
\newblock {\em Phys.\ Lett.}, 26B:8, {(1967)}.

\bibitem{1968BO16}
{J. P. Bondorf, P. von Brentano, P. Rich\-ard}.
\newblock {\em Phys.\ Lett.}, 27B:5, {(1968)}.

\bibitem{1968CR05}
{J. G. Cramer, P. von Brentano, G. W. Phillips, H. Ejiri, S. M. Ferguson, W. J.
  Braithwaite}.
\newblock {\em Phys.\ Rev.\ Lett.}, 21:297, {(1968)}.

\bibitem{1968VO02}
{P. von Brentano, W. K. Dawson, C. F. Moore, P. Rich\-ard, W. Whar\-ton, H.
  Wieman}.
\newblock {\em Phys.\ Lett.}, 26B:666, {(1968)}.

\bibitem{1968WH02}
{W. R. Whar\-ton, P. von Brentano, W. K. Dawson, P. Rich\-ard}.
\newblock {\em Phys.\ Rev.}, 176:1424, {(1968)}.

\bibitem{Zai1968}
{S. A. A. Zaidi, L. J. Parish, J. G. Kulleck, C. F. Moore, P. von Brentano}.
\newblock {\em {Phys.\ Rev.}}, {165}:{1312}, {(1968)}.

\bibitem{1969RI10}
{P. Rich\-ard, P. von Brentano, H. Wieman, W. Whar\-ton, \\ W.G.Weitkamp,
  W.W.McDonald, D.Spalding}.
\newblock {\em Phys.\ Rev.}, 183:1007, {(1969)}.

\bibitem{1970KU13}
{J. G. Kulleck, P. Rich\-ard, D. Burch, C. F. Moore, W. R. Whar\-ton, P. von
  Brentano}.
\newblock {\em Phys.\ Rev.}, C2:1491, {(1970)}.

\bibitem{LMUrep2000p70}
{R. Hertenberger, Y. Eisermann, H.-F. Wirth, and G. Graw}.
\newblock {Maier-Leibnitz Laboratorium, annual report}.
\newblock {\em {Universit\"at M\"unchen}}, page~{70}, {(2000)}.
\newblock {http://www.bl.physik.uni-muenchen.de/bl\_rep/\\ jb2000/p70.ps}.

\bibitem{LMUrep2000p71}
{H.-F. Wirth, H. Angerer, T. von Egidy, Y. Eisermann, G.Graw, and
  R.Hertenberger}.
\newblock {Maier-Leibnitz Laboratorium, annual report}.
\newblock {\em {Universit\"at M\"unchen}}, page~{71}, {(2000)}.
\newblock {http://www.bl.physik.uni-muenchen.de/bl\_rep/\\ jb2000/p71.ps}.

\bibitem{Wirth1999}
{H.-F. Wirth}.
\newblock {Ph.D. thesis}.
\newblock {\em {}}, {(2001)}.
\newblock {http://tumb1.biblio.\\ tu-muenchen.de/publ/diss/ph/2001/wirth.html}.

\bibitem{RalfsQ2005}
{R. Hertenberger, A. Metz, Y. Eisermann, K. El Abiary, A. Ludewig, C. Pertl, S.
  Trieb, H.-F. Wirth, P. Schiemenz, G. Graw,}.
\newblock {\em {Nucl.\ Inst.}}, {A536}:{266}, {(2005)}.

\bibitem{AB1973}
{A. Heusler and P. von Brentano}.
\newblock {\em {Ann.\ Phys.\ (N.Y.)}}, {75}:{381}, {(1973)}.

\bibitem{Mai1983}
{G. Mairle, K. Schindler, P. Grabmayr, G.J. Wagner, I. Schmidt-Rohr}.
\newblock {\em {Phys.\ Lett.}}, {121B}:{307}, {(1983)}.

\bibitem{Rej1999}
{M. Rejmund, M. Schramm, K. H. Maier}.
\newblock {\em {Phys.\ Rev.}}, {C59}:{2520}, {(1999)}.

\bibitem{BM1969}
{A. Bohr and Ben R. Mottelson}.
\newblock {\em {Nuclear Structure}}, volume~I.
\newblock {W. A. Benjamin, New York, Amsterdam}, {(1969)}.

\bibitem{Heu1969}
{A. Heusler, H. L. Harney and J. P. Wurm}.
\newblock {\em {Nucl.\ Phys.}}, {A135}:{591}, {(1969)}.

\bibitem{1971CL02}
{R. G. Clarkson, P. von Brentano, and H. L. Harney}.
\newblock {\em {Nucl.\ Phys.}}, {A161}:{49}, {(1971)}.

\bibitem{1952BB}
{J. M. Blatt, L. C. Biedenharn, and M. E. Rose}.
\newblock {\em {Revs.\ Mod.\ Phys.}}, {24}:{249}, {(1952)}.

\bibitem{Edm1957}
{A. R. Edmonds}.
\newblock {\em {Angular momentum in quantum mechanics}}.
\newblock {Princeton University Press}, {(1960)}.

\bibitem{AHwwwHOME}
{A. Heusler}.
\newblock {\em {http://www.mpi-hd.mpg.de/$\sim$hsl/}}.

\bibitem{Heu1969a}
{A. Heusler, M. Endriss, C. F. Moore, E. Grosse, P. von Brentano}.
\newblock {\em {Z.\ Physik}}, {227}:{55}, {(1969)}.

\bibitem{Latz1979}
{G. Latzel and H. Paetz Gen. Schieck}.
\newblock {\em {Nucl.\ Phys.}}, {A323}:{413}, {(1979)}.

\bibitem{Rie2005}
{F. Riess}.
\newblock {\em {http:\-www.cip.\-physik.\-uni-muenchen.\-de/$\sim$riess/}}.

\bibitem{LMUrep2004}
{A. Heusler, G. Graw, R. Hertenberger, H.-F. Wirth, P. von Brentano}.
\newblock {Maier-Leibnitz Laboratorium, annual report}.
\newblock {\em {Universit\"at M\"unchen}}, page~{21}, {(2004)}.
\newblock {http://www.bl.physik.uni-muenchen.de/bl\_rep/\\ jb2004/p021.ps}.

\bibitem{LMUrep2003}
{A. Heusler, G. Graw, R. Hertenberger, H.-F. Wirth, P. von Brentano}.
\newblock {Maier-Leibnitz Laboratorium, annual report}.
\newblock {\em {Universit\"at M\"unchen}}, page~{21}, {(2003)}.
\newblock {http://www.bl.physik.uni-muenchen.de/bl\_rep/\\ jb2004/p21.ps}.

\bibitem{GNPH140CeIII}
{A. Heusler}.
\newblock {\em {Nucl.\ Phys.}}, {A141}:{667}, {(1970)}.

\bibitem{HJG1972}
{H.-J. Gl\"ockner}.
\newblock Master's thesis, {Universit\"at Heidelberg}, {(1972)}.

\end{thebibliography}
\end{document}